\documentclass[openacc]{rsproca_new}

\titlehead{Research}

\begin{document}

\title{A mathematical framework for dynamic emergent constraints in climate science}

\author{
Francesco Ragone$^{1}$, Valerio Lucarini$^{1,2}$}

\address{$^{1}$School of Computing and Mathematical Sciences, University of Leicester, UK.\\
$^{2}$School of Sciences, Great Bay University, Dongguan, P.R. China.}

\subject{Earth Sciences, Statistical Physics, Mathematical Modelling}

\keywords{Emergent Constraints, Climate Change, Response Theory, Causality}

\corres{Francesco Ragone\\
\email{fr120@leicester.ac.uk}}

\begin{abstract}
Emergent constraints in climate science are empirical relations that link the response to a forcing of a physical observable to the properties of other observables, with the aim of reducing climate change projection uncertainties. Here we use recent results in linear response theory to develop a  mathematical framework for dynamic emergent constraints, a class of emergent constraints linking the response of different observables to the same forcing. 

We show how traditional dynamic emergent constraints are a special case of more general relations, that we call integral dynamic emergent constraints. These relations allow to compute the response of a predictand as the convolution of the response of a predictor and the proxy Green’s function of the predictand-predictor pair. The conditions for the existence of integral emergent constraints are related to the causality of the proxy Green’s function and the time scales at which the system is observed. We apply this framework to global warming simulations with the MPI-ESM climate model, to study dynamic emergent constraints between different observables. 

These results allow to put the theory of dynamic emergent constraints on firm mathematical ground, and suggest a protocol to identify necessary conditions for the existence of such relations in climate data.
\end{abstract}


\maketitle


\section{Introduction}

Climate projections are unavoidably characterized by uncertainties that affect projections across a range of spatial and temporal scales. Whilst uncertainties tend to become larger as we narrow our focus to smaller and smaller scales, nontrivial uncertainties exist also when considering global indicators like the Equilibrium Climate Sensitivity \cite{vonderHeydt2016,IPCC_2021_WGI}. Uncertainties are typically classified in three categories \cite{Nowack2025}. Scenario uncertainties are due to the lack of knowledge of which future greenhouse gases scenario will actually occur. Model uncertainties are due to errors in representing physical processes in climate models. Finally, sampling uncertainties are due to the chaotic nature of the climate system, which means that the forced response signal is intermixed with internal variability acting on a wide spectrum of time scales \cite{GhilLucarini2020}. Whilst scenario uncertainties can be addressed by explicitly considering different emission scenarios, model and sampling uncertainties need to be taken into account at the level of individual climate change experiments.

Emergent constraints have been proposed as a way to reduce projection uncertainties \cite{Klein2015,Cox2018,Hall2019} and consist of empirical relations between the response to a forcing of a given physical observable (the predictand) and the properties of some other physical observable (the predictor). The goal is to extract information from a better constrained or more accurately known quantity (the predictor) on a more uncertain one (the predictand). These empirical relations are typically identified with linear regressions on populations of different climate models or observational datasets. The idea is that if the properties of observable A are better represented in models than those of observable B, and an emergent constraint is found between the two, then observable A can be used as a predictor to constrain the projected response of observable B in future global warming scenarios. Examples of applications include emergent constraints on Equilibrium Climate Sensitivity \cite{Knutti2006,Cox2018,Nijsse2020,Williamson2021}, radiation  \cite{VAROTSOS2025106556}, precipitation \cite{Shiogama2022}, cloud properties \cite{Gordon2014}, biogeochemical indicators \cite{Wenzel2014}, and cryosphere and Arctic climate response \cite{Hall2006,Qu2014,Bracegirdle2012,TerhaaR2020}.


Emergent constraints are by construction statistics-based, empirical relations - more recently supplemented by machine learning approaches \cite{Nowack2025} - which are in some cases informed by the physical understanding of the climate processes involved. A fundamental classification divides emergent constraints into static and dynamic ones \cite{Nijsse2018,Williamson2021}. Static emergent constraints are (typically linear) relations between the response of the predictand in a forcing scenario and the statistical properties of the predictor in an unperturbed reference state. Dynamic emergent constraints are instead relations (also typically linear) between the response of the predictand and the response of the predictor to the same or a similar forcing, and involve a time dimension.

Despite these fundamental differences, it has been attempted to formalize both types of emergent constraints using tools borrowed from linear response theory \cite{Ruelle1998,Ruelle2009,Bettolo2008,HairerMajda2010,Sarracino2019,Santos2022,Lucarinietal2026}. Static emergent constraints have been justified forms of fluctuation-dissipation relations \cite{Cox2018}, as part of a long history of attempts at applying the fluctuation-dissipation theorem to climate data following the landmark contribution by Leith \cite{Leith1975}; see \cite{GhilLucarini2020} for an overview. Dynamic emergent constraints have been linked to a special form of linear response theory for systems of stochastic differential equations by \cite{Nijsse2018}. Despite the clear merit of these studies, a complete theory of emergent constraints able to a) define under which conditions a specific predictand and a specific predictor can be paired, and b) provide a mathematical expression of the link between the predictor and the predictand, is still lacking.

A promising approach to the problem comes from recent results in linear response theory that try to define causal pathways within a forced system. The key result obtained is the derivation of so-called proxy response formulas  \cite{Lucarini2018,Tomasini2021} , which relate the responses of different observables of a system undergoing an external forcing. The idea is to investigate under which conditions one can take one of the observables (the predictor) as a proxy of the acting forcing, and develop response formulas for predicting the response of a second observable (the predictand) based on the response of the first one. The goal is to be able to study the change in a system without necessarily having full information on the external forcing, identifying good predictors that surrogate efficiently the action of the external forcing on a range of predictands.


In this paper we use the results of \cite{Lucarini2018,Tomasini2021} to develop a mathematical formulation of dynamic emergent constraints, generalising the analysis of \cite{Nijsse2018}. We formalize the concept of dynamic emergent constraint in the context of proxy linear response theory, showing how traditional dynamic emergent constraints are a particular case of a broader class of relations, that we call integral dynamic emergent constraints. We then apply the theory to simulations with the coarse resolution (CR) version of the Max Planck Institute Earth System Model (MPI-ESM) v.1.24 \cite{Mauritsen2019}. We show how the existence of dynamic emergent constraints between pairs of physical observables depends on the causality of the corresponding proxy response function and on the coarse graining time scale at which the system is observed. The observables we consider are the globally averaged surface temperature, the global precipitation (as well as its large-scale and convective components), and the intensity of the Atlantic Meridional Overturning Circulation (AMOC). Finally we discuss how our findings can be interpreted in the context of different notions of causality \cite{Granger1969,Pearl2009}.

The rest of the paper is structured as follows. In Section \ref{sec:response_general} we present the theoretical results. In Section \ref{sec:results} we analyse global warming experiments with the MPI-ESM model, using the same setup used in \cite{Lembo2020}. Finally in Section \ref{sec:conclusions} we present our conclusions and discuss future research directions. In the Appendix \ref{Appa} we present in greater detail the numerical procedures used to process the data.

\section{\label{sec:response_general} Response and causality}

\subsection{\label{sec:dynamic_emergent_constraints}Dynamic emergent constraints}

Following the Hasselmann paradigm \cite{hasselmann1976,imkeller2001,LucariniChekroun2023}, we represent the dynamics of the unperturbed climate system as a stochastic dynamical system of the form $\dot{x}=F(x) +\sigma \eta(x,t)$, where $x$ is the state variable describing the large-scale, slow variables of the system, $F(x)$ is the drift and  defines the deterministic contribution to the dynamics, whilst $\eta(x,t)$ defines the noise law describing the stochastic forcing. The stochastic term provides a surrogate description of the impact of the fast, small scales of motions on the large scale, slow ones; see a technical discussion in \cite{imkeller2001,LucariniChekroun2023}. A climate change scenario can be represented by introducing a non-autonomous component $\gamma(x,t)$ to the dynamics, so that the perturbed evolution equation can be written as $\dot{x}=F(x) +\epsilon \gamma(x,t)+\sigma \eta(x,t)$. The forcing $\gamma(x,t)=B(x)f(t)$ depends on a function $B(x)$, which defines its structure in phase space, and a time modulation $f(t)$, whilst $\epsilon$ is a scaling constant. The function $B(x)$ determines the physical nature of forcing. For instance, the increase in greenhouse gases concentrations, or changes in aerosol loading, or a change in an orbital parameter, will all determine different functional forms for $B(x)$; see also discussion in \cite{BodaiLucarini2020Chaos}. The scaling constant $\epsilon$ and the time modulation $f(t)$ determine instead the type of scenario. For example, in the case of greenhouse gases emissions these could represent the different IPCC Representative Concentration Pathways (RCP) \cite{IPCC_2021_WGI}.

We consider physical observables as functions of the state of the system $\Phi(t)=\Phi(x(t))$. This could be for example the global average of surface temperature, or a locally defined observable as temperature at specific grid points \cite{Ragone2016,Lucarinietal2017}.  We indicate the expectation value of an observable in the unperturbed state ($\epsilon = 0$) as  $\langle \Phi\rangle _0$. Practically speaking, this corresponds to the average of the observable taken over many  members of an ensemble simulation initialised from independent initial conditions, or a time average taken over a long time. When we apply a forcing to the system starting from the unperturbed state, the expectation value of the observable will change and become time dependent. We indicate the time dependent expectation value under a given forcing as $\langle \Phi\rangle_{\epsilon f}(t)$. Note that in this paper we will consider the physical nature of the forcing fixed, so that, in order to simplify the notation, we do not indicate the explicit dependence on $B(x)$. The quantity $\langle\Phi_{\epsilon f}\rangle(t)$  can be estimated by taking the average over many ensemble members, where each member undergoes the same forcing \cite{Bodai2013,Lucarinietal2017,Tel2020}, just as done in the case of CMIP simulations \cite{Eyring2016}, of when constructing large ensemble simulations \cite{Maher2021}. Instead, time averaging cannot be used here because the statistical properties of the system depend explicitly on time. 

The response of an observable to a forcing scenario is the difference $\delta\Phi_{{\epsilon f}}(t)=\langle\Phi\rangle_{\epsilon f}(t)-\langle\Phi\rangle_0$. Now let us consider two observables, a predictand $\Phi_{1}(t)$ and a predictor $\Phi_{2}(t)$.  A linear dynamic emergent constraint can be expressed mathematically as the existence of a constant $\alpha_{\Phi_1\Phi_2}$ such that for every time scenario $\epsilon f(t)$ the following relation holds
\begin{equation}
\delta\Phi_{1,{\epsilon f}}(t)=\alpha_{\Phi_1\Phi_2}\delta\Phi_{2,{\epsilon f}}(t)
\label{dec}
\end{equation}
with $\alpha_{\Phi_1\Phi_2}$ independent of $\epsilon f(t)$ \cite{Nijsse2018,Williamson2021}. We remind that the response of the observables here is considered in expectation value, which means that Eq. \ref{dec} is consistent with the usual practice of identifying emergent constraints with linear regressions. Note that, whilst Eq. \ref{dec} may seem fairly restrictive, it can accommodate for a wide range of physical relations. For any non-linear relation of the form $\delta\Phi_{1,{\epsilon f}}(t)=r(\delta\Phi_{2,{\epsilon f}}(t))$ it is always possible to find a suitable invertible transformation of one of the two observables that leads to a linear relation, as long as $r$ is monotonic. Consequently, any non-linear but monotonous relation that holds across different scenarios can be expressed as a dynamic emergent constraint of the form given in Eq. \ref{dec}.

\subsection{\label{sec:proxy_response}Proxy linear response}

Response theory states that the expectation value of an observable under the action of a forcing in the form introduced above can be written as 
\begin{equation}
  \langle \Phi\rangle_{\epsilon f}(t)=\langle\Phi_{0}\rangle+\sum_{n=1}^{+\infty}\epsilon^n\Phi_{f}^{(n)}(t).
\end{equation}    
The sum on the right hand side gives the response  $\delta\Phi_{{\epsilon f}}(t)$ expressed as a power series expansion in $\epsilon$ \cite{ruelle_nonequilibrium_1998,Lucarini2008}. The first term of the series gives the linear response and can be computed as
\begin{equation}
\Phi_{f}^{(1)}(t)=\intop_{-\infty}^{+\infty}G_{\Phi}(t-s)f(s)\textrm{d}s,
\label{response}
\end{equation}
where $G_{\Phi}(t)$ is the linear Green's function of the observable. The Green's function is in general a causal function, i.e. $G_{\Phi}(t)=0$ for $t<0$. The integral in Eq. \ref{response} can therefore be taken up to time $t$ \cite{Ruelle1998,Ruelle2009}. Physically, this means that the information of the forcing propagates onto physical observable only forward in time. The Green's function depends on the forcing type $B(x)$, but not on the scenario $\epsilon f(t)$. Consequently, if the Green's function of an observable for a given forcing type is known, it is possible to compute the linear response of the observable to any scenario $\epsilon f(t)$ \cite{Ragone2016,Lembo2020}. The Green's function can be estimated by performing a suitable set of probe experiments where a specific choice of the time modulation $f(t)$ , as described in  \cite{Ragone2016,Lucarinietal2017} and in the following. Alternatively, it can be estimated by making use of different variants of the fluctuation-dissipation theorem  \cite{Abramov2007,Bettolo2008,Cooper2011}, by taking advantage of machine learning methods \cite{giorgini2024linear,giorgini2024datadriven}, or by combining response theory with Koopman theory \cite{Lucarini2025,Zaglietal2026,Lucarinietal2026}. 


The Fourier transform \cite{Arfken2013} of the linear Green's function is the linear susceptibility $\chi_{\Phi}(\omega)$, which gives the linear response at angular frequency $\omega$ as
\begin{equation}
\hat{\Phi}_{f}^{(1)}(\omega)=\chi_{\Phi}(\omega)\hat{f}(\omega).
\end{equation}
where $\hat{\Phi}_{f}^{(1)}(\omega)$ and $\hat{f}(\omega)$ are  the Fourier transforms of ${\Phi}_{f}^{(1)}(t)$  and $f(t)$ respectively. The fact that $G_{\Phi}(t)$ is a causal function implies that $\chi_{\Phi}(\omega)$ admits analytic continuation in the upper complex plane \cite{Titchmarsh1939}, and that its real and imaginary parts are a pair of Hilbert transforms, leading to the Kramers-Kronig dispersion relations \cite{Ruelle1998,Lucarini2008}. Since $\chi_{\Phi}(\omega)$ is the Fourier transform of the causal function $G_{\Phi}(t)$,  its analytic continuation $\chi_{\Phi}(\sigma)$ is the Laplace transform \cite{Arfken2013} of $G_{\Phi}(t)$
\begin{equation}
\chi_{\Phi}(\sigma)=\chi_{\Phi}(\lambda+i \omega)=\int_0^{+\infty} G_{\Phi}(t)e^{-(\lambda+i\omega) t}\textrm{d}t,
\end{equation}
where $\sigma=\lambda+i \omega$ the complex frequency with rate $\lambda$ and angular frequency $\omega$. 

Linear response operators can be used to reconstruct the linear response to a forcing of an observable using the linear response of another observable to the same forcing \cite{Lucarini2018, Tomasini2021}. Let us consider a predictand $\Phi_1(t)$ and a predictor $\Phi_2(t)$. Using Eq. 4 we have 
$\hat{\Phi}_{1,f}^{(1)}(\omega)=\chi_{\Phi_1}(\omega)\hat{f}(\omega)$ and $\hat{\Phi}_{2,f}^{(1)}(\omega)=\chi_{\Phi_2}(\omega)\hat{f}(\omega)$, where $f(t)$ is the same in both equations. Therefore we have
\begin{equation}
\hat{\Phi}_{1,f}^{(1)}(\omega)=\frac{\chi_{\Phi_1}(\omega)}{\chi_{\Phi_2}(\omega)}\hat{\Phi}_{2,f}^{(1)}(\omega)=\chi_{{\Phi_1\Phi_2}}(\omega)\hat{\Phi}_{2,f}^{(1)}(\omega),
\end{equation}
where 
\begin{equation}
\chi_{{\Phi_1\Phi_2}}(\omega)=\frac{\chi_{\Phi_1}(\omega)}{\chi_{\Phi_2}(\omega)}\label{proxy_susceptibility},
\end{equation}
is the proxy susceptibility of the predictor-predictand pair \cite{Lucarini2018, Tomasini2021}. Taking the inverse Fourier transform of both sides of the equation we have the proxy response formula
\begin{equation}
\Phi_{1,f}^{(1)}(t)=\intop_{-\infty}^{+\infty}G_{\Phi_1\Phi_2}(s)\Phi_{2,f}^{(1)}(t-s)\textrm{d}s,
\label{proxy_response}
\end{equation}
where $G_{\Phi_1\Phi_2}(t)$ is the proxy Green's function of the predictor-predictand pair, computed as the inverse Fourier transform of $\chi_{{\Phi_1\Phi_2}}(\omega)$. Using Eq. \ref{proxy_response} one can therefore reconstruct the response of a predictand from the response of a predictor, provided that the proxy Green's function of the predictor-predictand pair is known and that the response of the predictor has been observed for all time $t$ on the entire real axis. Note also that in general  $G_{\Phi_1\Phi_2}(t)\neq G_{\Phi_2\Phi_1}(t)$, as the relation between the two observable will be in general asymmetric.

\subsection{\label{sec:causality} Integral emergent constraints and causality}

Differently from a regular Green's function, a proxy Green's function in general will not be a causal function. Let $S_{\Phi_1\Phi_2}$ be the set of complex frequencies $\sigma$ in the upper complex plane at which the analytic continuation $\chi_{\Phi_{2}}(\sigma)$ of the predictor susceptibility  has complex zeros, whilst the analytic continuation  $\chi_{\Phi_1}(\sigma)$ of the predictand susceptibility does not. Their ratio $\chi_{\Phi_1}(\sigma)/\chi_{\Phi_{2}}(\sigma)$ will thus have singularities for $\sigma \in S_{\Phi_1\Phi_2}$. Since the analytic continuation of the ratio of two functions, if it exists, is the ratio of the analytic continuations of the two functions, a sufficient and necessary condition for $\chi_{\Phi_1\Phi_2}(\omega)$ to admit analytic continuation is that $S_{\Phi_1\Phi_2}$ is empty. Then its analytic continuation $\chi_{\Phi_1\Phi_2}(\sigma)$ exists and it is equal to the ratio 
\begin{equation}
\chi_{\Phi_1\Phi_2}(\sigma)=\frac{\chi_{\Phi_1}(\sigma)}{\chi_{\Phi_{2}}(\sigma)}\label{proxy_susceptibility_continuation},
\end{equation}
and $G_{\Phi_1\Phi_2}(t)$ is a causal function. If instead $S_{\Phi_1\Phi_2}$ is not empty, the proxy susceptibility $\chi_{\Phi_1\Phi_2}(\omega)$ does not admit analytic continuation, and its inverse Fourier transform $G_{\Phi_1\Phi_2}(t)$ is not a causal function. Note that in general  $S_{\Phi_1\Phi_2}\neq S_{\Phi_2\Phi_1}$, which implies again that in terms of causality  the relationship between predictor and predictand is in general asymmetric. 

If the proxy Green's function is a causal function, the integral in \ref{proxy_response} can be taken starting from $t=0$, as the proxy Green's function is zero for negative time lag. Assuming a forcing   $f(t)=0$ for $t<0$, then also $\Phi_{2,f}^{(1)}(t)=0$ for $t<0$, and therefore equation \ref{proxy_response} becomes
\begin{equation}
\Phi_{1,f}^{(1)}(t)=\intop_{0}^{t}G_{\Phi_1\Phi_2}(s)\Phi_{2,f}^{(1)}(t-s)\textrm{d}s.
\label{proxy_response_causal}
\end{equation}
This result is more powerful than Eq. \ref{proxy_response}, since it only requires the response of predictor for time instants in the past of the lead time $t$, thanks to the causality of the proxy Green's function. Therefore, observing the response of the predictor on a finite time domain $[0,T]$ (as it happens in practice) allows to compute the response of the predictand on the the same interval $[0,T]$ from the knowledge of the proxy Green's function.

Equation \ref{proxy_response_causal} provides a generalisation of the concept of dynamic emergent constraint that takes into account not only the state of the predictor at the time $t$ when the predictand is evaluated, but also its history starting with the activation of the forcing up to time $t$. The classic definition of a  dynamic emergent constraint $\Phi_{1,f}^{(1)}(t)=\alpha_{\Phi_1\Phi_2}\Phi_{2,f}^{(1)}(t)$ is recovered when the proxy Green's function can be approximated as a Dirac delta 
\begin{equation}
G_{\Phi_1\Phi_2}(t)\approx\alpha_{\Phi_1\Phi_2}\delta(t)
\label{proxy_response_causal_delta}
\end{equation}
for a given constant $\alpha_{\Phi_1\Phi_2}$. This happens when the time scale at which the signals are observed is much larger than the time scale of decay of the proxy Green's function.

In the following we will refer to the  general case \ref{proxy_response_causal} as integral dynamic emergent constraint, and to the special case \ref{dec} as instantaneous dynamic emergent constraint. The integral form of the more general result means that the past history of the predictor is able to surrogate all the information necessary to determine the change in the properties of the predictand at a later time. In this case, the predictor can be interpreted as acting as a reaction coordinate for the system, i.e. a quantity that controls its macroscopic properties \cite{Zaglietal2026}. Note also that, given the conditions discussed above, a predictor will be good regardless of the choice of the predictand, except for the case of special degeneracies. This might explain the reason why so many so-called spurious emergent constraints have been found in the literature, whereby spurious refers to the fact that no obvious physical link is ascertained between the predictor and the predictand. 

If the proxy Green's function instead is non-causal, in order to reconstruct the response of the predictand it is necessary to take the integral  in equation \ref{proxy_response}  on the entire real axis, including negative time lags, which correspond to the effect of future values of the predictor on the present values of the predictand. This means that if one observes the responses on a finite time domain $[0,T]$, the response of the predictand can not be fully reconstructed from the response of the predictor, as this would require knowledge of the response of the predictor at unobserved (future) time outside of $[0,T]$. In such a situation, dynamic emergent constraints are not possible, neither in integral nor in instantaneous form. As we will show in the following,  filtering the signal by performing temporal coarse-graining of the response of the observables to focus only on certain time scales (a common practice of data processing in climate science) can enforce causality of the proxy Green's function and thus the emergence of dynamic constraints. 

We can introduce a quantitative measure of the degree of non-causality of a proxy Green's function introducing a causality index which measures how close a proxy Green's function is to a causal function. Let $G_{\Phi_1\Phi_2}^{+}(t)$ and $G_{\Phi_1\Phi_2}^{-}(t)$ be respectively the causal and non-causal parts of $G_{\Phi_1\Phi_2}(t)$, that is $G_{\Phi_1\Phi_2}^{+}(t)=G_{\Phi_1\Phi_2}(t)$ for $t\ge0$ and $G_{\Phi_1\Phi_2}^{+}(t)=0$ for $t<0$, whilst $G_{\Phi_1\Phi_2}^{-}(t)=G_{\Phi_1\Phi_2}(t)$ for $t<0$ and $G_{\Phi_1\Phi_2}^{-}(t)=0$ for $t\ge0$. We define the causality index $C_{\Phi_1\Phi_2}$ as 
\begin{equation}
    C_{\Phi_1\Phi_2}=1-\frac{||G_{\Phi_1\Phi_2}^{-}||}{||G_{\Phi_1\Phi_2}^{+}||+||G_{\Phi_1\Phi_2}^{s}||}\label{causalindex}
\end{equation} 
where $||\cdot ||$ indicates the $L_2$ measure, and $G_{\Phi_1\Phi_2}^{s}(t)$ is the singular component of $G_{\Phi_1\Phi_2}(t)$, a singular term that can can appear at $t=0$ as discussed in \cite{Lucarini2018, Tomasini2021}. In practical applications where we deal with discrete data this term is simply considered part of $G_{\Phi_1\Phi_2}^{+}(t)$ at $t=0$ \cite{Lucarini2018, Tomasini2021}. Note that Eq. \ref{causalindex} is slightly different from the definition in \cite{Lucarini2018, Tomasini2021}. With this definition, $C_{\Phi_1\Phi_2}$ is 1 for a perfectly causal function and 0 for a perfectly non-causal function, and an integral dynamic emergent constraint will thus emerge for $C_{\Phi_1\Phi_2}$ close to 1. As usual, in general $C_{\Phi_1\Phi_2}\neq C_{\Phi_2\Phi_1}$ due to the asymmetry of the causal relations between the two observables.

\section{\label{sec:results}Results}

\subsection{Data}

We apply the framework of proxy response described above to simulations with the coarse resolution (CR) version of the Max Planck Institute Earth System Model (MPI-ESM) v.1.24 \cite{Mauritsen2019}. The setup is the same used in \cite{Lembo2020}.The model includes as atmospheric module ECHAM673 with T31 spectral resolution (equivalent to 96 gridpoints in longitude and 48 in latitude) and 31 vertical levels. The oceanic module is MPI-OM74 on a curvilinear orthogonal bipolar grid (GR30) (with 122 longitudinal and 101 latitudinal gridpoints) with 40 vertical levels. See \cite{Lembo2020} for more details on the setup of the model. 

We perform one control run and two forcing scenarios. The control run is a 2000 years long stationary run in pre-industrial conditions with $CO_2$ concentration set at 280 ppm. The first forcing scenario ($H_2$) is an ensemble simulation with 20 ensemble members, each 1910 years long and initialized from a different initial condition taken from the control run at intervals of 100 years, with $CO_2$ concentration doubled abruptly at the beginning of each run. The second forcing scenario ($R_2$) is set up in the same way, but the simulations are 1000 years long and the $CO_2$ concentration is increased by 1$\%$ per year until it has doubled with respect to the pre-industrial value (after about 70 years), and is kept constant afterwards. We refer to this as ramp experiment in the following. These are standard forcing scenarios in climate modelling, used to calculate key response metrics like the Equilibrium Climate Sensitivity and the Transient Climate Response respectively \cite{IPCC_2021_WGI}.

Going back to the functional description of the perturbation $\epsilon B(x)f(t)$, in both experiments the forcing  features the same term $B(x)$. From a physical point of view this term projects on the thermodynamic variables of the atmosphere and modulates the radiative effect of $CO_2$. Hence, the forcing is implemented as a change in the input of the parametrization of the radiative transfer module of the climate model. Since in both experiments the $CO_2$ concentration stabilizes at the same level, it is natural to consider that for both experiments the scaling constant is the same, and we denote it as $\epsilon_{2\times CO_2}$. The two forcing then differ only in the time modulation $f(t)$. For the $H_2$ experiment $f(t)=H(t)$, where $H(t)$ is the Heaviside function, with $H(t)=0$ for $t<0$ and $H(t)=1$ for $t \ge 0$. Instead, for the $R_2$ experiments the forcing can be taken as a linear function until the time of the doubling and constant afterwards
\begin{equation}
f(t)=\begin{cases}
\frac{t}{\tau}, & 0\le t < \tau \\
1 , & t \ge \tau
\end{cases}
\end{equation}
where $\tau=$70 years. The 1$\%$ per year increase in $CO_2$ concentration corresponds to a linear $f(t)$ modulating the forcing because to a very good degree of approximation the radiative forcing scales with the logarithm of the $CO_2$ concentration. This representation of the forcing has been successfully employed in applications of response theory to climate change simulations  \cite{Ragone2016,Lucarinietal2017,Lembo2020}.

With this experimental setup, it is then possible to derive the Green's function associated to an observable $\Phi(t)$ directly from the estimate of $\epsilon_{2\times CO_2}\Phi_{H_2}^{(1)}(t)$. This is computed under the assumption of linearity as $\epsilon_{2\times CO_2}\Phi_{H_2}^{(1)}(t)\approx \delta \Phi_{H_2}(t)= \langle\Phi\rangle_{H_2}(t)-\langle\Phi\rangle_{0}$, where $\Phi_{H_2}(t)$ is the ensemble average of the  signal of the perturbed ensemble to the $H_2$ forcing, and $\langle\Phi\rangle_{0}$ is estimated as the time average of $\Phi(t)$ in the control run. The linearity of the response has been tested for these simulations in \cite{Lembo2020}, and the procedure to compute the Green's function from $\epsilon_{2\times CO_2}\Phi_{H_2}^{(1)}(t)$ is detailed in \cite{Ragone2016} and in the Appendix. The $R_2$ experiment is then used as a validation set, to test the prediction of obtained with response theory against direct simulations. Note that formally neither $B(x)$ nor $\epsilon_{2\times CO_2}$ are determined explicitly, but they are not needed in order to apply the response formulas, as discussed in the Appendix.

We consider the following physical observables: the globally averaged annual near surface temperature $T_s$, the globally averaged annual precipitation rate $P$, and the annual average of the intensity of the Atlantic Meridional Overturning circulation $M$, which is defined as the vertically integrated mass weighted meridional mass streamfunction across lat 26.5$^o$ N in the Atlantic Ocean as in \cite{Lembo2020}. We also consider convective precipitation $P_c$ and large scale precipitation $P_l$ separately, where $P=P_c+P_l$. Convective and large scale precipitation are computed by different parameterization schemes in climate models due to the different dynamical  processes that generate them (whilst the microscopic processes are the same), and are the dominant contributions to total precipitation in tropical and mid-high latitude areas respectively \cite{Peixoto}.

\begin{figure*}
\includegraphics[scale=0.45]{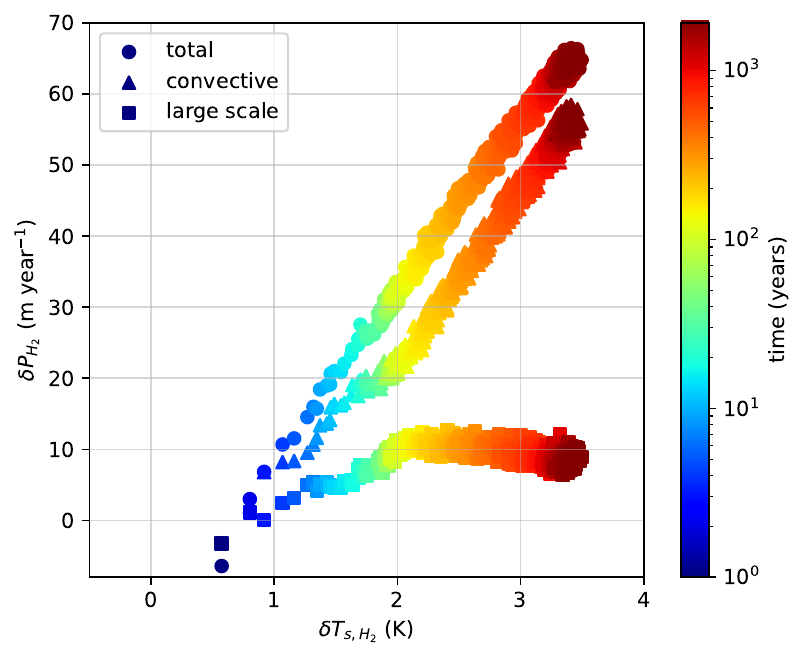}a
\includegraphics[scale=0.45]{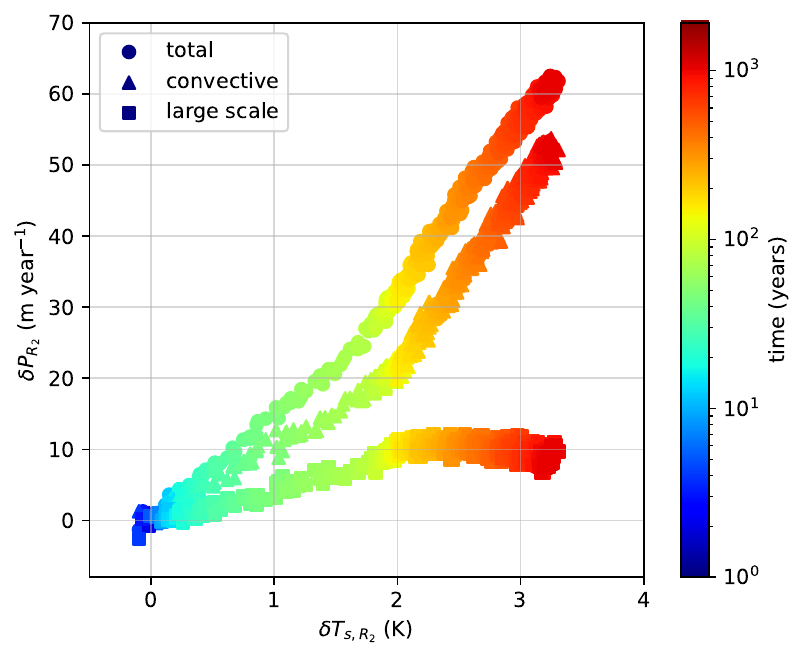}b
\includegraphics[scale=0.45]{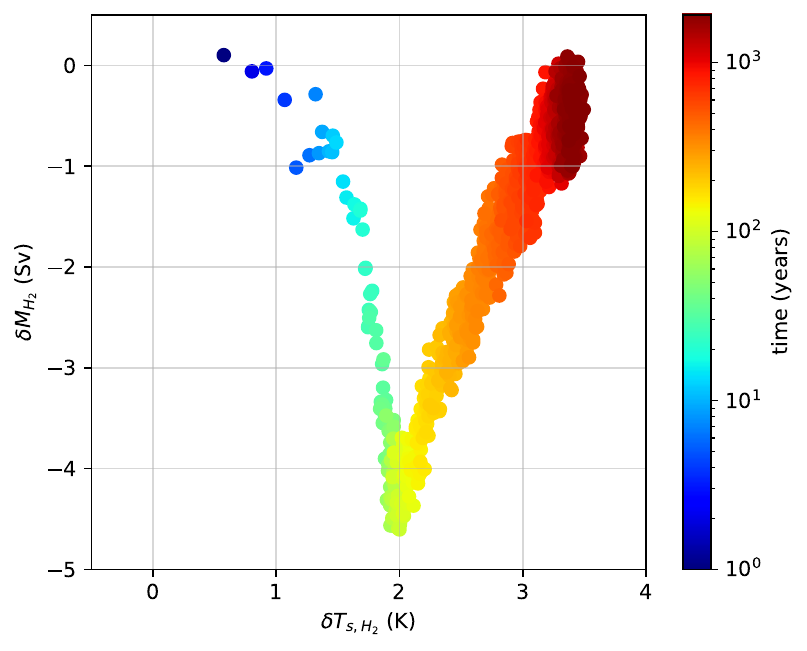}c
\hspace{23pt}\includegraphics[scale=0.45]{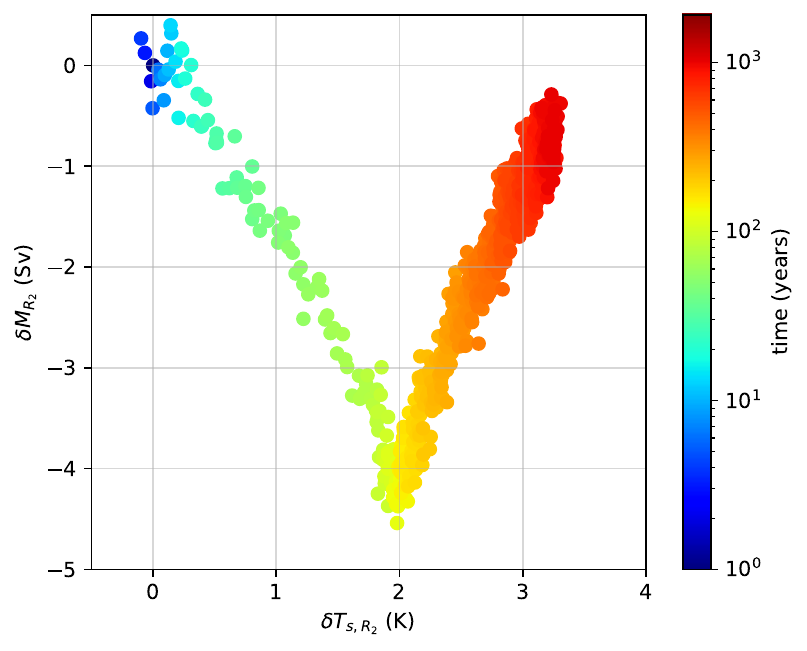}d
\caption{\label{fig:wide} Scatter plot of ensemble average response of global precipitation (total, convective and large scale) vs ensemble average response of global surface temperature to instantaneous $CO_2$ doubling (a) and to 1$\%$ py  $CO_2$ ramp increase (b). Scatter plot of ensemble average response of AMOC index vs ensemble average response of global surface temperature to instantaneous $CO_2$ doubling (c) and to 1\% py  $CO_2$ ramp (d). The colors show time in logarithmic scale.}\label{scatter}
\end{figure*}

\subsection{Response of physical observables}

In the $H_2$ experiment, the change of the total precipitation $P$ scales linearly with the change of the surface temperature $T_s$  (Figure \ref{scatter}a). This reflects the well known fact that there is a strong thermodynamic control on global total precipitation by the lower tropospheric temperature, of which surface temperature acts as a proxy \cite{HeldSoden2006}. The slope of $P$ as a function of $T_s$ is about 22 $m\, y^{-1} K^{-1}$, which corresponds to about a 2$\%$ increase per degree of warming. Consistently with most Earth System model simulations and projections \cite{Stephens2008}, this value is lower than the 7$\%$ increase per degree of warming expected from the Clausius-Clapeyron relation (CC) for the global moisture atmospheric content. Note that the instantaneous response of total precipitation is negative: immediately after the $CO_2$ doubling, during the first year $P$ has a sudden decrease, followed by the long term linear increase. This effect is due to the initial atmospheric warming caused by the direct effect of $CO_2$ increase, that leads to a decrease of the relative humidity of the atmosphere and therefore of global precipitation, which is then overcome as soon as the sea surface temperature rises and more moisture is taken up by the atmosphere \cite{Zappa2020}.

Looking separately at the convective and at the large scale precipitation, a more complex picture emerges (Figure \ref{scatter}a). Both $P_c$ and $P_l$ increase nonlinearly for the first 100 years, until the temperature response reaches 2$K$. At this value, a regime change occurs, where $P_l$ stops increasing and starts a very slow decrease. In both regimes there is a compensation between $P_c$ and $P_l$ that leads $P$ to increase linearly at the same constant rate, even after the regime shift. This suggests that the thermodynamic control by temperature acts on global scale, whilst the different behavior of the individual components is attributable to changes in dynamics resulting from  global warming. In particular, the stall in the increase of the large scale precipitation is attributable to compensating effect of the  decrease in the baroclinicity of the climate system and in the increase of moisture in the mid-latitude atmosphere \cite{HaerteR2009,Berg2013,Pendergrass2014}. 

In the $R_2$ experiment, up to the stabilization of the $CO_2$ concentration there is a linear relationship between the change in $T_s$ and the change in $P_c$ and $P_S$ (Figure 1b). Yet, the link between these quantities is different from the previous case, which indicates that the link between the temporal paths of change is non-trivial. It is not enough to know by how much  temperature has changed at a given time, to be able to say by how much the precipitation has changed at that time. In other terms, an equation like Eq. \ref{proxy_response_causal_delta} cannot hold true in this case. After the stabilization of the $CO_2$ concentration, the relationship between the observables mimics closely what found in the $H_2$ scenario. Note that the fact that immediately after reaching doubled $CO_2$ the response behaves exactly like in the $CO_2$ doubling case is not as easily visible if one looks only at the time series of the individual observables \cite{Lembo2020}, and a robust structure is visible only in the relation among variables, not in the time evolutions taken individually.

The regime shift in precipitation response is associated to the recovery of the AMOC strength $M$ after an initial decrease (Figure 1c). In the $H_2$ scenario, the $M$ decreases very rapidly until the 2$^o$ $K$ global warming threshold, and then recovers very slowly to almost the control value, linearly with $T_s$ to a good degree of approximation. The recovery of the AMOC after a sudden decrease is a common feature of abrupt $CO_2$ increase scenarios in state of the art climate models  \cite{Nobre2023}. This is due to the negative feedbacks of the ocean circulation that eventually erode the stratification in the deep water formation regions caused by the changes in the freshwater forcing and temperature anomalies associated with global warming. In the $R_2$ scenario, the relationship between $ M$ and $ T_s$ is different from the $H_2$ case during the ramp period (Figure 1d), which  indicates that the link between AMOC intensity and global temperature change again is non-trivial, as in the case the precipitation.

These two cases showcase radically different behaviours. The relation between the response of temperature and precipitation is monotonic (with the partial exception of large scale precipitation), while the relation between the response of temperature and AMOC is non-monotonic. It seems intuitive that the presence of monotonicity between two climate observables suggests the possibility of establishing an emergent constraint between the two, whilst non-monotonicity suggests the opposite. As we show below, these expectations are not necessarily met in an obvious way once the general integral formulation discussed above is adopted and the skill of prediction is evaluated for different time scales.

\begin{figure*}
\includegraphics[scale=0.4]{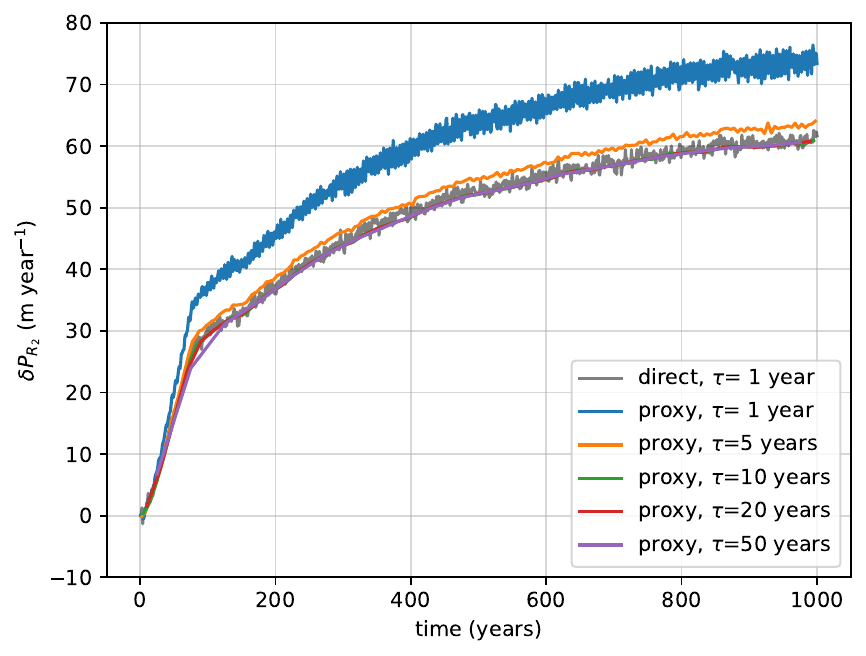}a
\includegraphics[scale=0.4]{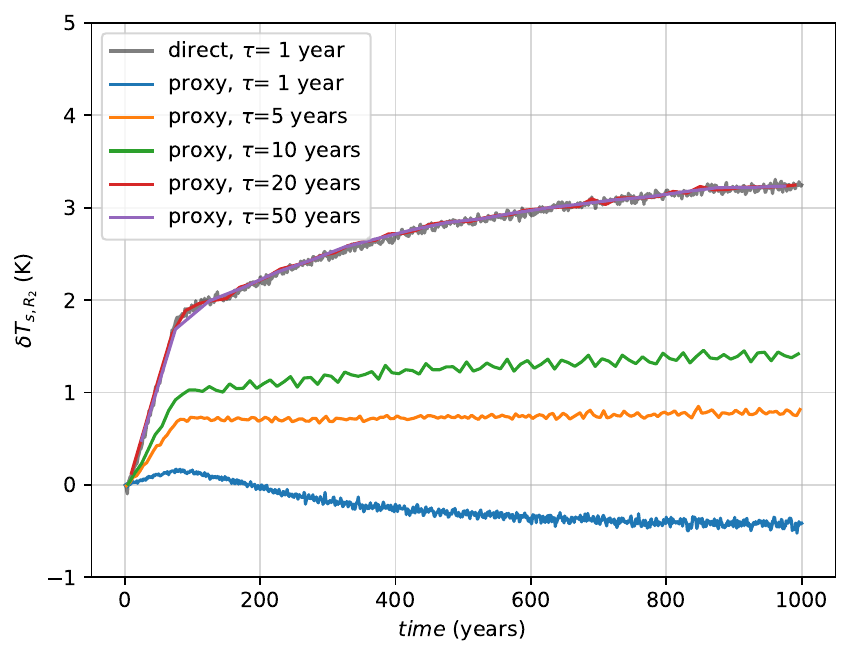}b
\includegraphics[scale=0.4]{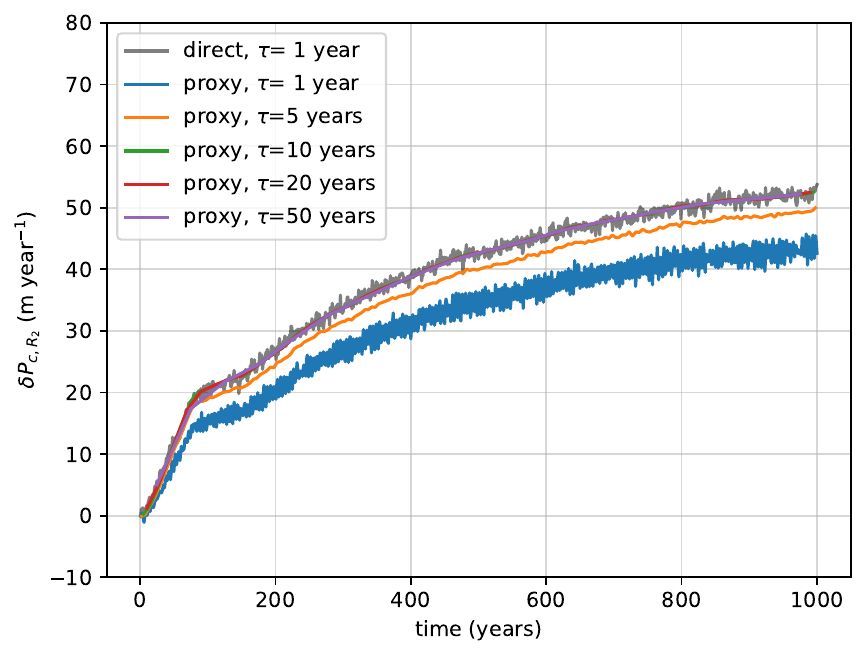}c
\includegraphics[scale=0.4]{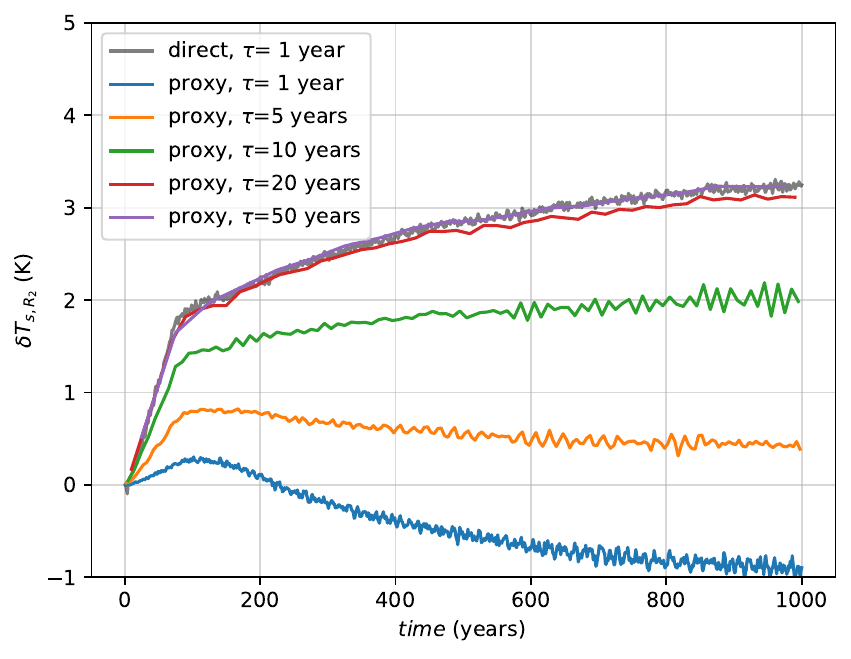}d
\includegraphics[scale=0.4]{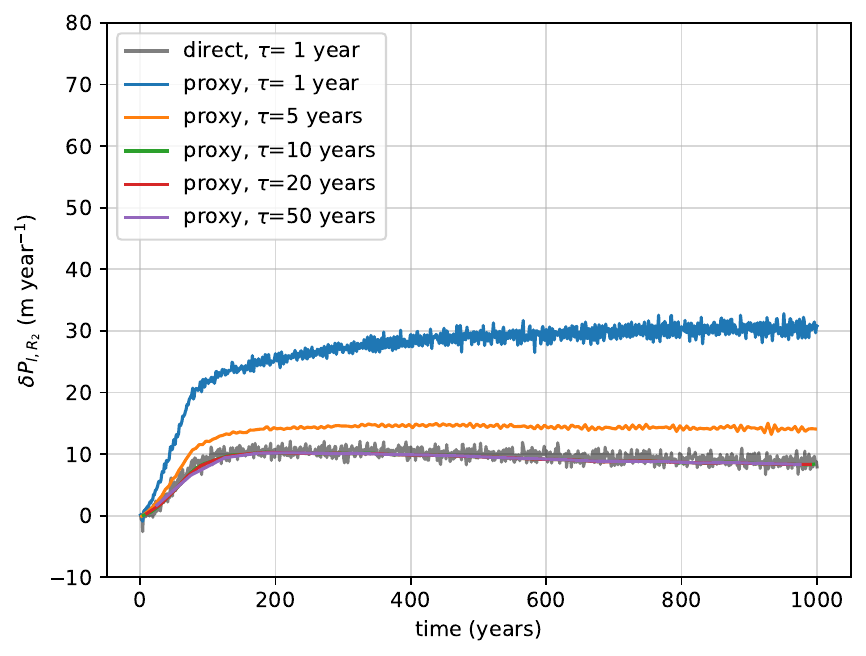}e
\hspace{29pt}\includegraphics[scale=0.4]{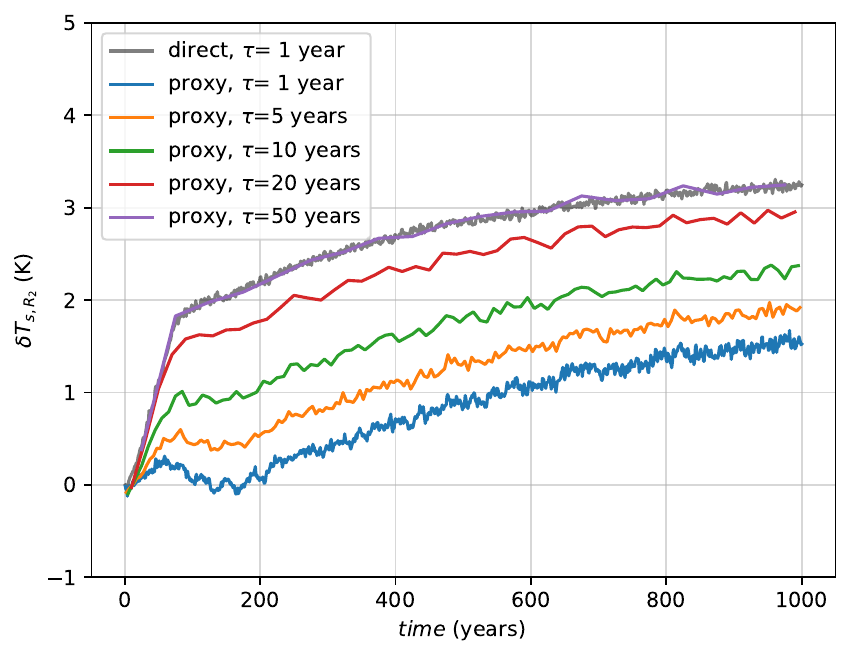}f
\caption{\label{fig:wide} Global surface temperature response, original direct simulation at one year average (grey) and proxy reconstruction using global (a), convective (c) and large scale (e) precipitation as predictor, for different temporal coarse graining (colours). Global (b), convective (d) and large scale (f) precipitation response, original direct simulation at one year average (grey) and proxy reconstruction using global surface temperature as predictor, for different temporal coarse graining (colours). }\label{PvsT}
\end{figure*}

\subsection{Proxy response analysis}

To test the framework described in the previous section, we use the results of the $H_2$ experiments to compute proxy Green's functions for pairs of observables at different temporal coarse-graining levels. The coarse-grained time series of the response of an observable $\Phi(t)$ is computed as
\begin{equation}
      {\overline  \Phi_{\tau}(t_j)}=\frac{1}{\tau}\int_{j\tau}^{(j+1)\tau}\Phi(s)\mathrm{d}s
\end{equation} 
where $\tau$ is the coarse-graining time scale and  $t_j=j\tau$, with $j$ an integer index. We consider different values for the coarse graining time scale $\tau$, ranging from 1 year (the original temporal resolution of our data) up to 80 years. For each coarse graining time scale, the corresponding  coarse grained proxy Green's functions are computed applying the procedure described in the Appendix on the coarse-grained time series of the predictor-predictand pairs.  

We then compute proxy response predictions for the $R_2$ experiment by applying equation \ref{proxy_response_causal} for a given predictand $\Phi_1(t)$ using the observed response in the $R_2$ experiment of a predictor $\Phi_2(t)$ and the proxy Green's functions for that pair of observables computed from the $H_2$ experiment.  We stress that we use formula \ref{proxy_response_causal}, which means that we take the integral ranging from  0 to $t$. This means that if the proxy Green's function includes a non-causal part, we expect the proxy response prediction to fail to reconstruct the actual response. 

\subsubsection{Global surface temperature and precipitation}

We first consider the case of  $T_s$ and $P$. When using $P$ as predictand and $T_s$ as predictor (Figure \ref{PvsT}a), the proxy response for $\tau=1$  largely overestimates the actual response. This disagreement suggests that the proxy Green's function has a large non-causal component. Increasing the coarse graining time scale, however, the prediction converges to the actual response. At a coarse graining time scale of 10 years and longer the prediction is basically perfect, indicating that at these time scales the proxy Green's function is causal. 

If we take $T_s$ as predictand and $P$ as predictor (Figure \ref{PvsT}b) we can see that the reconstruction at annual time scale is completely wrong, as the predicted $T_s$ even decreases with $P$, and the monotonicity between the quantities is lost. Increasing the coarse graining time scale the proxy response prediction converges to the actual response. However, in this case it is necessary to take a larger coarse graining time scale, at least around 20 years, to have a good match. Therefore there is a range of time scales $\tau$, between 10 and 20 years, where global surface temperature  can be used as a causal predictor of global total precipitation in a integral dynamic emergent constraint, but not the other way around. Convective and large scale precipitation show a similar picture (Figures  \ref{PvsT}c, \ref{PvsT}d\ref{PvsT}e, and \ref{PvsT}f). The most notable difference is that the convergence to a correct reconstruction of the response of $T_s$ is slower for $P_l$ than for $P$ and $P_c$.

\begin{figure*}
\includegraphics[scale=0.4]{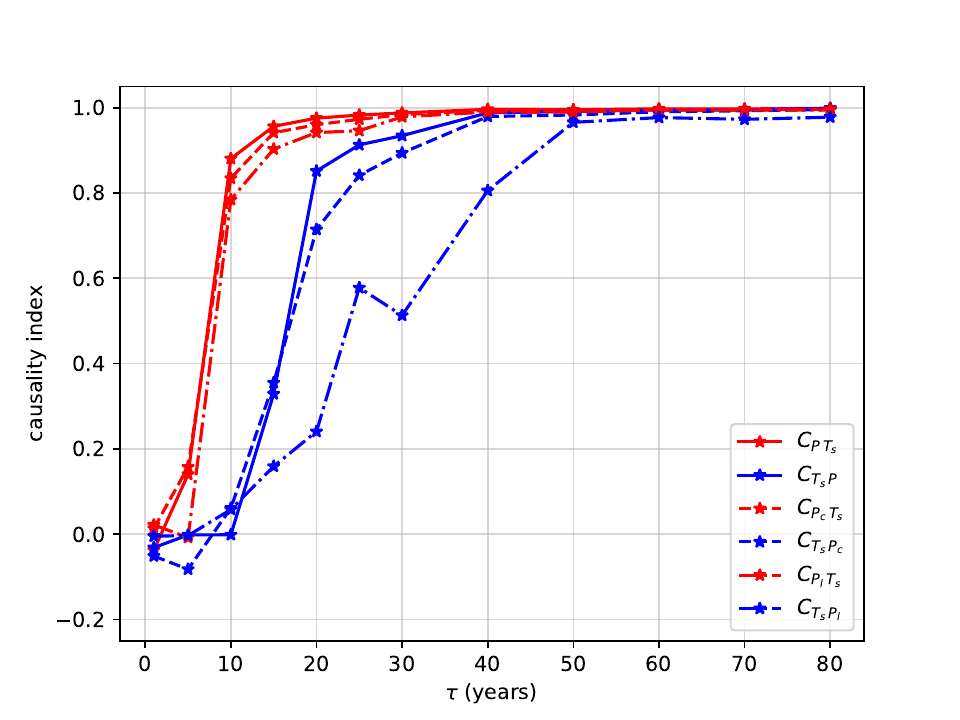}a
\includegraphics[scale=0.4]{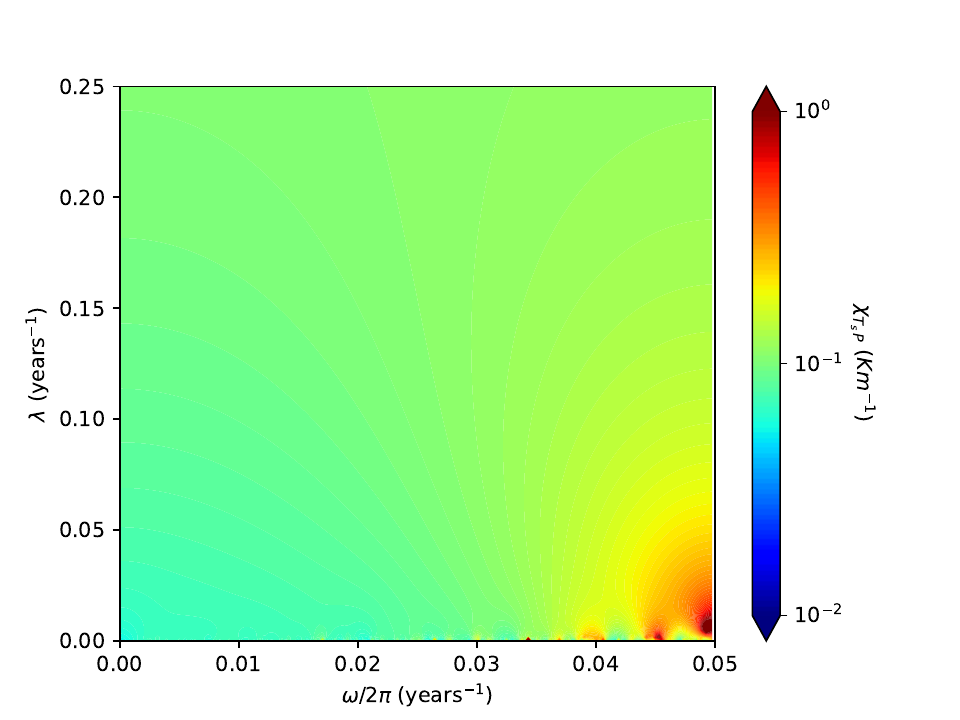}b
\includegraphics[scale=0.4]{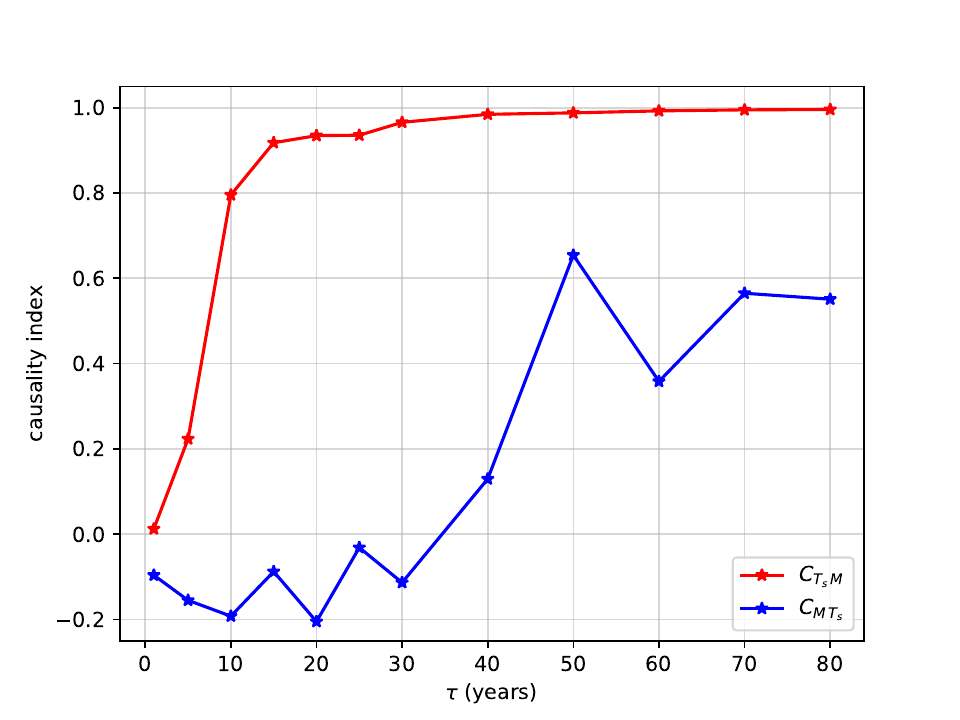}c
\includegraphics[scale=0.4]{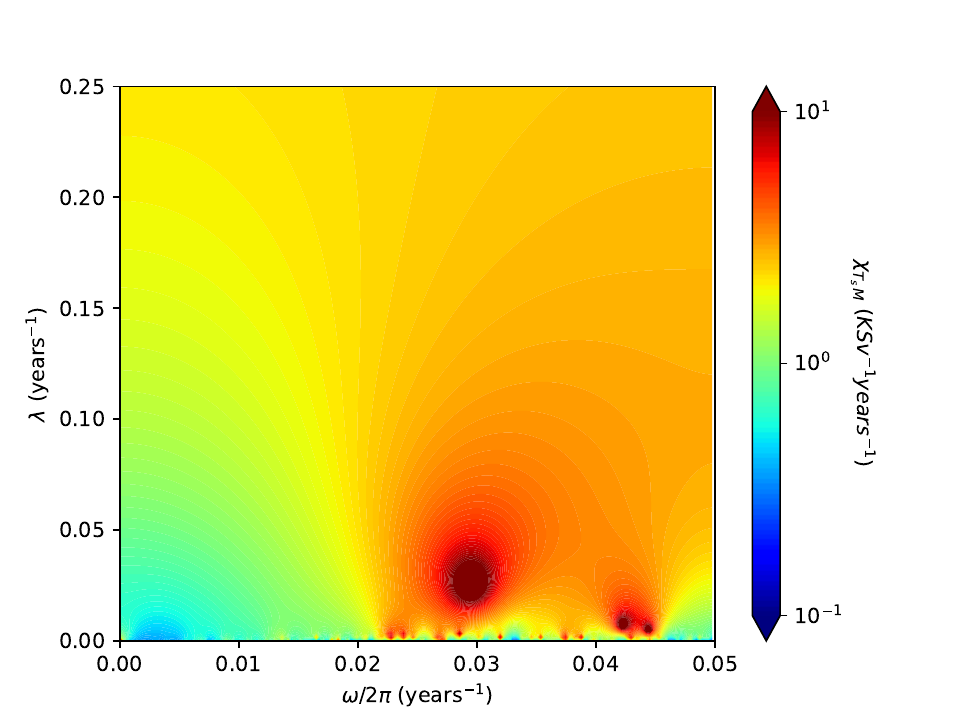}d
\caption{\label{fig:wide}a) Causality index as function of coarse graining time scale $\tau$, in red (blue) global surface temperature as predictor (predictand) and global total, convective and large scale precipitation as predictand (predictor). b) Module of proxy susceptibility with global total precipitation as predictor and global surface temperature as predictand, as function of imaginary (horizontal axis) and real (vertical axis) component of complex frequency. c) Causality index as function of coarse graining time scale $\tau$, in red (blue) global surface temperature as predictor (predictand) and AMOC index as predictand (predictor). d) Module of proxy susceptibility with AMOC index as predictor and global surface temperature as predictand, as function of imaginary (horizontal axis) and real (vertical axis) component of complex frequency.}\label{causality_index}
\end{figure*}

The lack of skill in the reconstruction of the response is due to the presence of a non-causal component in the proxy Green's function. In order to quantify this, in Figure \ref{causality_index}a we show the causality index $C_{\Phi_1\Phi_2}$ defined in Section \ref{sec:causality} as a function of the coarse graining time scale $\tau$, for the six cases discussed above. When using $T_s$ as predictor, for both $P$, $P_c$ and $P_l$ the causality index starts at 0 at annual time scale, but it increases quite rapidly and reaches a plateau between 10 and 20 years coarse graining time scale. At these time scales the causality index is above 0.9, and reaches approximately 1 when we consider $\tau\geq30$ years. When using precipitation as a predictor for $T_s$, we observe that the convergence is generally slower with increasing $\tau$: the causality index reaches 0.9 at around 25 years time scale for $P$ and $P_c$, slightly faster for $P$ than for $P_c$. In order to reach this value for $P_l$, we need to consider $\tau\geq50$ years. 

Physically it makes sense that large scale precipitation response is less informative on surface temperature response than convective precipitation response. Whilst the formation of convective precipitation is related to the atmospheric vertical lapse rate, which is directly influenced by surface temperature increase, large scale precipitation is triggered by baroclinic instability, that has a more complex dynamical relation to surface temperature. The fact that $P$ is the best predictor is also not surprising, because the thermodynamic constraint linking temperature and moisture impacts precipitation as a whole.  

This analysis shows how the causality relation between the response of global surface temperature and the response of global precipitation depends critically on the coarse-graining time scale $\tau$ at which the system is observed. At annual time scale there is virtually no relation, and the response of one is not informative to infer the response of the other. At decadal time scale the response of global surface temperature is a good descriptor of the response of global precipitation and its components, but not the other way around. This suggests that at this time scale global precipitation response is mainly driven by global surface temperature response, and it can be parametrized by it with an integral dynamic constraint. The fact that the converse is not true implies that the integral form of the constraint is irreducible, and that the pathway of change of surface temperature up to lead time $t$ matters to determine precipitation response at time $t$. 

At multidecadal time scale (20 years and larger) an integral dynamic emergent constraint exists in both directions. This implies that the response of the two observables contains essentially the same information, and that an instantaneous dynamic constraint can exist. It is interesting to notice that this time scale is corresponds to the one (usually 30 years) that is traditionally used to define the climatology  at operational level (e.g. according to WMO standard), when the focus is on traditional meteorological variables like, precisely, surface temperature and precipitation.


\subsubsection{Causality and time coarse-graining}

These results suggest that causal relations (in the sense of proxy response) among the forced responses of different observables depend on the time scale at which these observables are observed. This can be explained by the theory developed in the previous Section. Let us assume that ${\sigma}_{\Phi_1\Phi_2}={\lambda}_{\Phi_1\Phi_2}+i{\omega}_{\Phi_1\Phi_2}$ is a complex frequency in $S_{\Phi_1\Phi_2}$ such that $\chi_{\Phi_1\Phi_2}({\sigma})$ has a singularity at $\sigma_{\Phi_1\Phi_2}$ and no singularities for all the complex frequencies $\sigma=\lambda+i\omega$ such that $\omega < \omega_{\Phi_1\Phi_2}$.  Let us assume that the data we have access to are coarse-grained on a time scale $\tau$. This could mean that our data are averages over $\tau$ (for example, annual or decadal averages) or that they are sampled on a sampling period $\tau$  (for example, paleoclimatic data sampled every century). The susceptibilities estimated from the observed response signals will be sampled up to a maximum frequency $\omega_{max}$ such that $\omega_{max}/2\pi=1/2\tau$. Let us define the time scale $\tau_{\Phi_1\Phi_2}$ such that if the data were coarse-grained on  $\tau_{\Phi_1\Phi_2}$ the corresponding cut-off frequency would be $\omega_{\Phi_1\Phi_2}$, that is  $\omega_{\Phi_1\Phi_2}/2\pi=1/2\tau_{\Phi_1\Phi_2}$. If $\tau>\tau_{\Phi_1\Phi_2}$, then $\omega_{max}<\omega_{\Phi_1\Phi_2}$ and no singularities will be spectrally resolved in the proxy susceptibility. In this case the corresponding proxy Green's function $G_{\Phi_1\Phi_2}(t)$ will therefore be a causal function. On the contrary, if $\tau\le{\tau}_{\Phi_1\Phi_2}$ then some singularities will be resolved, and $G_{\Phi_1\Phi_2}(t)$ will include a non-causal component. The same reasoning can be done in the opposite direction, giving a critical coarse-graining time scale $\tau_{\Phi_2\Phi_1}$ for having a causal $G_{\Phi_2\Phi_1}(t)$.

To illustrate this analysis, in Figure \ref{causality_index}b we show the module of  ${\chi}^{test}_{T_sP}=\chi_{T_s}(\sigma)/\chi_{P}(\sigma)$, with $ \sigma=\lambda +i\omega$, as a function of $\omega/2\pi$ and $\lambda$, for a coarse graining time scale $\tau$=10 years. We can see that there are a number of isolated local maxima for low values of $\lambda$ and high values of $\omega$. These sharp local maxima indicate the presence of complex zeroes of $\chi_{P}(\sigma)$ not matched by complex zeros of $\chi_{T_s}(\sigma)$. The presence of the non-matched zeroes implies that ${\chi}_{T_sP}(\sigma)$ does not exist as analytic continuation of ${\chi}_{T_sP}(\omega)$,  and  that  $G_{T_sP}(t)$ is not a causal function. Note that when dealing with finite data and therefore a discrete frequency domain, these do not appear as true singularities, but rather as sharp local maxima at the sampled frequency closer to the true singularity (see the Appendix). The non-causality of $G_{PT_s}(t)$  would appear instead as the presence of localized minima in Figure  \ref{causality_index}b. The presence of local maxima and the absence of local minima is consistent with the fact that at $\tau=10$ years the causality index $C_{PT_s}$ is close to 1, whilst the causality index $C_{T_sP}$ is very low, close to zero. Taking a coarse graining on a time scale of 25 years is equivalent to filter out all frequencies larger than 0.02 years$^{-1}$, and thus all the singularities. Consequently, also $C_{T_sP}$ in this case would increase to nearly 1, consistently with Figure  \ref{causality_index}a.

\begin{figure}
\includegraphics[scale=0.4]{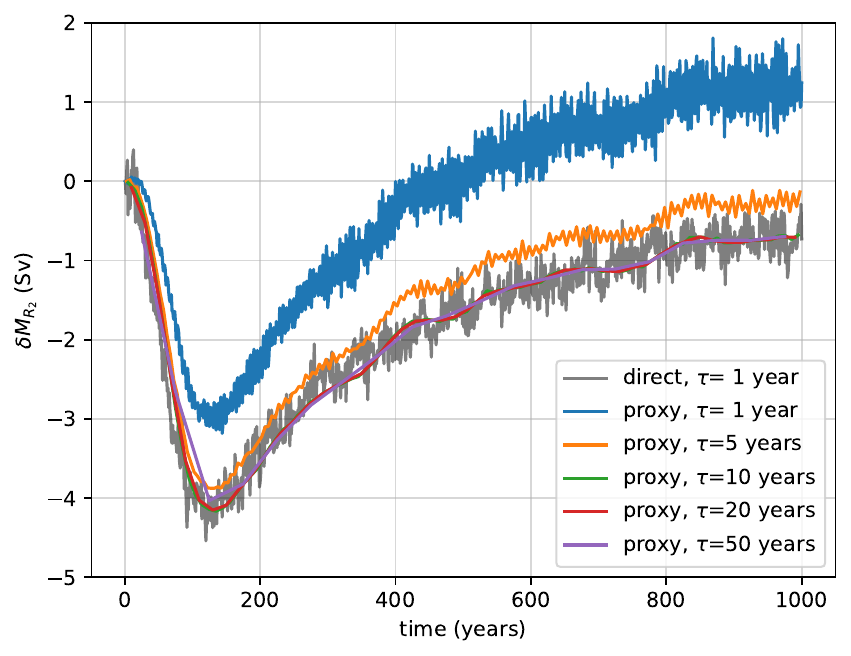}a
\includegraphics[scale=0.4]{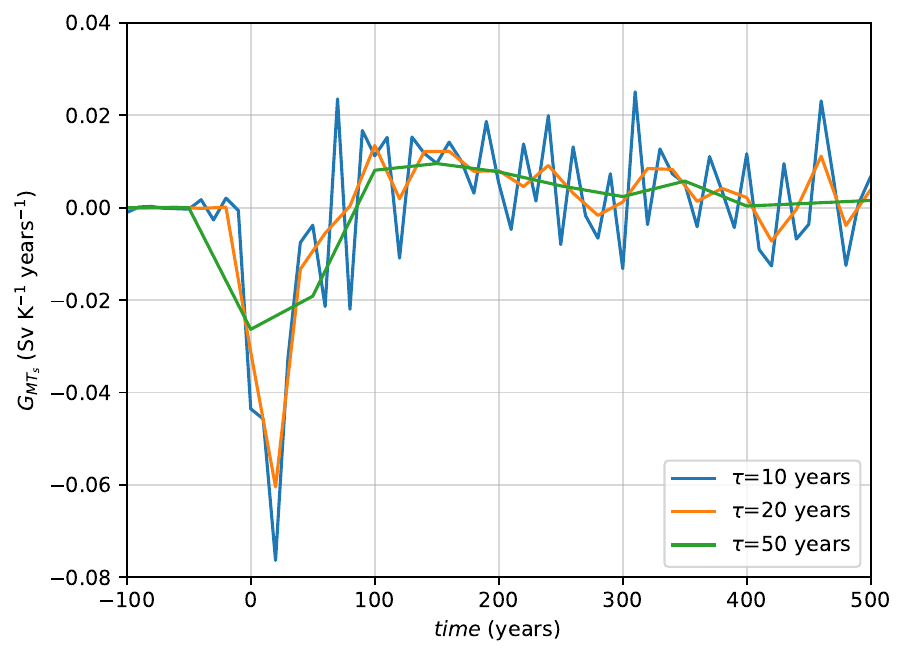}b
\includegraphics[scale=0.4]{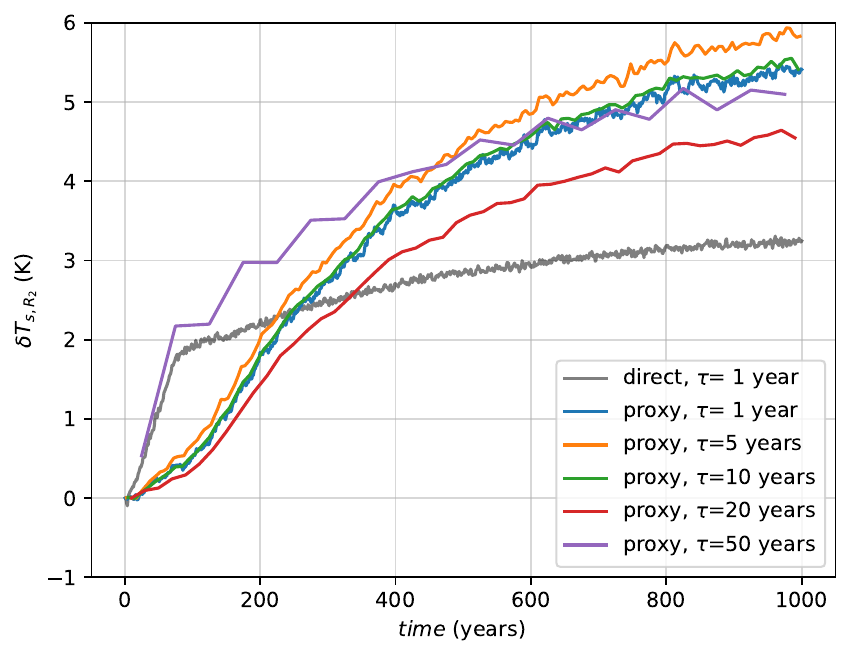}c
\hspace{23pt}\includegraphics[scale=0.4]{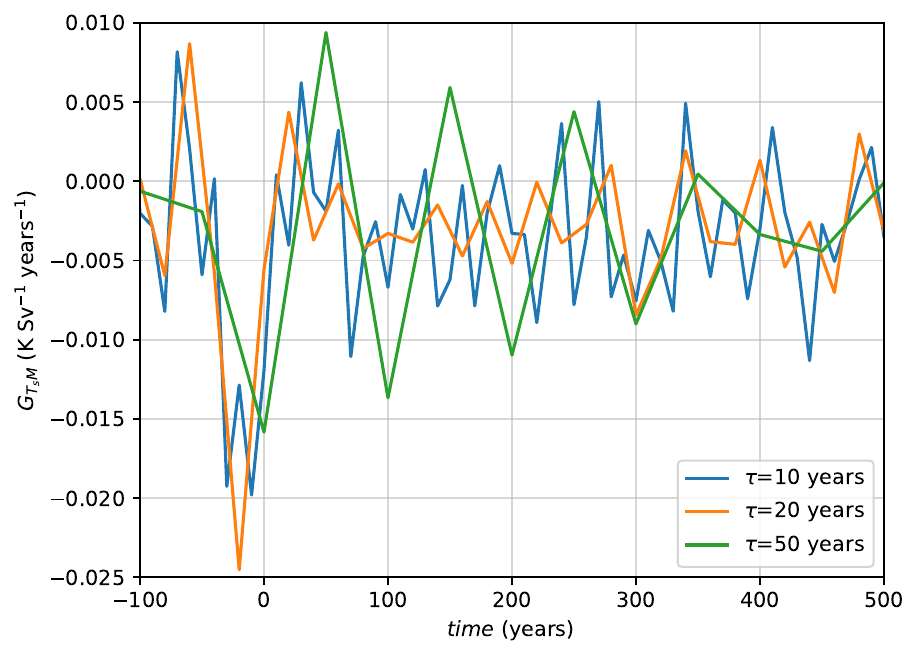}d
\caption{\label{fig:AMOC_response}a) AMOC index response, observed annual averages (grey) and proxy reconstructed using global surface temperature as predictor for different temporal coarse graining (colours). b) Proxy Green's function with global surface temperature as predictor and AMOC index as predictand for different temporal coarse graining. c) Global surface temperature response, observed annual averages (grey) and proxy reconstructed using AMOC index as predictor for different temporal coarse graining (colours). d) Proxy Green's function with AMOC index as predictor and  global surface temperature as predictand for different temporal coarse graining. }
\end{figure}

\subsubsection{Global surface temperature and AMOC}

A different picture emerges when analysing the response of the AMOC. Figure \ref{fig:AMOC_response}a  show the response of the AMOC index $M$ and the proxy predictions obtained with $T_s$ as predictor for different coarse-graining time scales, while Figure \ref{fig:AMOC_response}c shows the same for $T_s$ as predictand and $M$ as predictor. Whilst the response of $T_s$ eventually converges to being a good predictor of the AMOC response at time scales larger than 10 years, the AMOC response is never a good predictor of the response of $T_s$, no matter how large we take the coarse graining time scale. Only when we consider $\tau=50$ y we are able to describe to some extent the initial $T_s$ response, yet the long term behaviour is completely off. 

This can be understood by looking at the corresponding proxy Green's functions in Figures \ref{fig:AMOC_response}b and \ref{fig:AMOC_response}d. The proxy Green's function for the case where $T_s$ is as predictor is causal for $\tau\ge$ 10 years. The proxy Green's function is negative for the first 100 years, and then becomes positive for longer time scales, decaying to zero at about 300-400 years. Since $T_s$ increases monotonically and considering the fact that prediction is performed via a convolution product, this is consistent with the fact that the AMOC  decreases for the first few centuries and then recovers on millennial time scales . The proxy Green's function for the case with the AMOC as predictor instead has an essential component at negative time lag that is present even at very large coarse graining time scales. Therefore the proxy Green's function is always non-causal in this case, and it is never able to causally reconstruct the $T_s$ signal. 

The causality index behaves as expected from the analysis of the reconstructions and of the proxy Green's functions (Figure \ref{causality_index}c). The causality index with $T_s$ as predictor converges rapidly to 1 at climatic time scales with $\tau$ larger than 10-20 years. On the contrary, the causality index with the AMOC as predictor is negative for $\tau$  up to 30-40 years, and then has a rapid increase and it reaches a plateau for $\omega_k/2\pi$ between 0.5 years $^{-1}$ and 0.6 years $^{-1}$ for $\tau$ larger than 50 years.  Correspondingly, the  module of  ${\chi}^{test}_{T_sM}=\chi_{T_s}(\sigma)/\chi_{M}(\sigma)$ shows a large local maximum at around $\omega/2\pi$=0.03 years$^{-1}$, and several smaller maxima for $\lambda$ nearly zero and $\omega/2\pi$ up to 0.015 years$^{-1}$. These singularities are filtered out taking $\tau$ larger than 30 years, consistently with the behavior of the causality index, that rapidly increases at those time scales. Nonetheless, the AMOC strength cannot be used as predictor of $T_s$ even when considering very long time scales. 

\section{\label{sec:conclusions} Conclusions}

In this paper we have shown how linear response theory allows to develop a robust mathematical framework for dynamic emergent constraints.  We have shown that under the assumption of linear response, one can obtain integral relations that express the response of a predictand to a forcing as the convolution between the response of a predictor to the same forcing and the proxy Green's function of the predictor-predictand pair. We called these relations integral dynamic emergent constraint, and we have discussed how they constitute a generalisation of the traditional concept of  dynamic emergent constraint \cite{Nijsse2018,Williamson2021}. The key difference between the two is that in the more general case, knowing the history of the predictor up the a time $t$ is necessary (and sufficient) to reconstruct the response of the predictand at time $t$.

That the validity of an integral dynamics emergent constraint requires the causality of the proxy Green's function. We have shown how for a given predictand-predictor pair $\Phi_1$ and $\Phi_2$, the causality of the proxy Green's function depends on which is the predictor and which is the predictand. This is in agreement with the intuition that different observables encode a more or less meaningful description of the overall dynamical processes, as  discussed in \cite{Tomasini2021}. This asymmetry, which is absent in the special case of traditional instantaneous dynamic emergent constraints, has been demonstrated analysing the proxy response of pair of observables in  the MPI-ESM climate model, where the proxy response functions have been computed from dedicated sets of experiments that correspond to standard practices in climate science.

Another key result is that the causality of the proxy Green's function and the validity of integral dynamic emergent constraints depend on the time scale at which the system is observed. In particular, there is in general a time scale  $\tau_{\Phi_1\Phi_2}$ that determines whether a coarse graining of the data on a time scale $\tau\geq\tau_{\Phi_1\Phi_2}$ is conducive to a causal proxy Green's function. Since due to the asymmetry usually $\tau_{\Phi_1\Phi_2}\neq \tau_{\Phi_2\Phi_1}$, assuming $\tau_{\Phi_1\Phi_2}<\tau_{\Phi_2\Phi_1}$ there can be three situations depending on  the coarse-graining time scale $\tau$:
\begin{enumerate}
\item If $\tau < \tau_{\Phi_1\Phi_2} < \tau_{\Phi_2\Phi_1}$ nether observable is a good causal predictor of the other one. No integral dynamic emergent constraint can exist in either direction, and therefore no instantaneous dynamic emergent constraint either.
\item If ${\tau}_{\Phi_1\Phi_2} \le \tau < {\tau}_{\Phi_2\Phi_1}$, the response of $\Phi_{2}$ can be used as a causal predictor of the response of $\Phi_{1}$, but not the other way around. An integral dynamic emergent constraint exists only with $\Phi_{2}$ as predictor and $\Phi_{1}$ as predictand. Also in this case an instantaneous dynamic emergent constraint cannot  be established .
\item If $\tau \ge \tau_{\Phi_2\Phi_1} >  \tau_{\Phi_1\Phi_2}$ each observable can be used as a causal predictor of the other. Integral dynamic emergent constraints exist in both directions. An instantaneous dynamic emergent constraint can exist for values of $\tau$ that are much longer than the time scales of all the feedbacks acting between $\Phi_1$ and $\Phi_2$. Such time scales control the decay of the proxy Green's functions, and the time scale $\tau$ required for full thermalization of the two observables will be in general longer than $\tau_{\Phi_1\Phi_2}$ and $\tau_{\Phi_2\Phi_1}$.
\end{enumerate}

These results show that the notion of emergent constraints to relate the response of different climate observables applies to many more cases than previously thought, provided that one uses the history of the observables up to the lead time in the general integral formulation, rather than seeking an instantaneous relation. This could in principle have important implications for the practice of proxy data reconstruction, where an instantaneous (usually or a relatively coarse-grained time scale) relationship between climatic variables is sought. We remark that integral relationship are linear, hence linear optimisation methods can be used to discover optimal relationship between predictands and predictors from data. 

The theory also indicates clear conditions for the validity of integral dynamic emergent constraints that have to be satisfied for such relations to hold. In other word, the methodology is not a silver bullet than can be used in all cases. This may also explain why in many cases emergent constraints simply do not apply. On the other hand, the theory predicts that good predictors are universally good for any choice of predictand. An interesting result of our analysis in this sense is that the globally averaged surface temperature is an effective predictor for coarse-graining of 10 years or more, which means that it acts as a good proxy of global warming at decadal scales. This indicates the somewhat reassuring fact that this quantity, that is traditionally taken as the main indicator of anthropogenic climate change, indeed surrogates effectively the impact of $CO_2$ increase on the global climate, because it controls to a first approximation the dynamics and the thermodynamics of climate at these scales. 

These results can also be related to classical notions of causality, although with some caveats. The non-causality of a proxy Green's function and the non-existence of an integral emergent constraint with observable $\Phi_1$ as predictand and observable $\Phi_2$ as predictor is an indication that the response of $\Phi_1$ cannot be a causal determinant of the response of $\Phi_2$. However, the causality of a proxy Green's function and the existence of an integral emergent constraint does not in general imply that the response of $\Phi_1$ is the cause of the response of $\Phi_2$. The response of $\Phi_1$ and the response of $\Phi_2$ are generally related by a complex network of feedback processes that involve a large number of hidden variables, triggered by a common external forcing, and disentangling causal information flows is a complex problem \cite{Pearl1995}. These aspects are also discussed  in \cite{Koutsoyiannis2022a, Koutsoyiannis2022b}, who obtained a formula similar to \ref{proxy_response_causal} using a different approach. Note that \cite{Koutsoyiannis2022a, Koutsoyiannis2022b} used  this argument to claim that $CO_2$ concentration increase is not the cause of global surface temperature increase. Ref. \cite{Asbrink2023} commented on this, showing that one can establish a causal link between $CO_2$ and global surface temperature change when multidecadal time scales are considered. The analysis presented in this paper provide the mathematical framework to explain those results. 

These aspects can be framed in terms of the difference between Granger and Pearl causality. If the proxy Green's function $G_{\Phi_1,\Phi_2}$ is causal, then $\Phi_1$ has a strong Granger  causality \cite{Granger1969} relationship with  $\Phi_2$. The case (ii) above indicates the situation where a clear asymmetry in the information flow is present between the two variables \cite{Allione2025,Deltatto2025}. However, the causality $G_{\Phi_1,\Phi_2}$ does not imply that there is a strong Pearl causality \cite{Pearl2009} relation between $\Phi_1$ and $\Phi_2$, because the change of $\Phi_1$ and $\Phi_2$ is due to the same common cause, that is the acting forcing; see a discussion on the link between Pearl causality and response theory in \cite{LucariniChekroun2024}. Another angle in this sense could be given by the Liang-Kleeman information flow theory \cite{Liang2016,PIRES2024133988}, whose results could be compared with our approach in a future work.

Finally, we stress again that the response functions depend on the function $B(x)$ that determines the physical nature of the forcing. This means, for example, that the response functions describing the linear response of global surface temperature to changes in the atmospheric $CO_2$ concentration will be different from that of the response to changes in an orbital parameter. In this paper we have only considered a forcing due to changes in $CO_2$ concentration. The  framework is however valid for any type of forcing, and, due to linearity, also for linear combinations of different forcing terms.

\vskip6pt

\enlargethispage{20pt}

\dataccess{The data from the model simulations are available at the World Data Center for Climate repository \cite{data2xCO2abrupt,data1pctCO2}. The code is available at \url{https://github.com/frragone/proxy_response}.}

\ack{FR and VL are grateful to J. Demaeyer, P. Cox, G. Zappa, R. Bastiaansen and V. Lembo for many useful exchanges on this topic. FR and VL acknowledge partial support by the ARIA SCOP-PR01-P003-Advancing Tipping Point Early Warning AdvanTip project. VL additionally acknowledges partial support provided by the Horizon Europe Projects Past2Future (Grant No. 101184070) and ClimTIP (Grant No. 100018693), by the European Space Agency Project PREDICT (Contract 4000146344/24/I-LR), and by the NSFC  International Collaboration Fund for Creative Research Teams (Grant No. W2541005).}


\appendix

\section{Data Analysis}\label{Appa}

\subsection{Computation of linear response functions from numerical experiments}

Green's functions and susceptibilities are computed from a set of experiments with the numerical climate model MPI-ESM \cite{Lembo2020} following the procedure described in \cite{Ragone2016}. We start from a 2000 years long control run  in preindustrial stationary conditions. We then consider an ensemble of 20 simulations where we abruptly double the $CO_2$ concentration ($H_2$ experiment). Each ensemble member is run for 1910 years, and the 20 initial conditions are taken at constant intervals of time from the control run (see \cite{Lembo2020} for more details). 

The abrupt forcing scenario can be represented as $H_2(t)=\epsilon_{2\times CO_2}H(t)$, where $H(t)$ is the Heaviside function with $t=0$ the time of the application of the forcing, and $\epsilon_{2\times CO_2}$ is a unknown scaling constant. Under the assumption of linearity, the response of an observable $\Phi (t)$ to the $H_2$ scenario is then
\begin{equation}
\delta\Phi_{H_2}(t)\approx\epsilon_{2\times CO_2}\Phi_{H}^{(1)}(t)=\intop_{-\infty}^{+\infty}\epsilon_{2\times CO_2}G_{\Phi}(t-s)H(s)\textrm{d}s=\intop_{0}^{t}\epsilon_{2\times CO_2}G_{\Phi}(s)\textrm{d}s.
\end{equation}
This allows to compute the Green's function up to $\epsilon_{2\times CO_2}$ as
\begin{equation}
\epsilon_{2\times CO_2}G_{\Phi}(t)\approx\frac{d}{dt}\delta \Phi_{H_2}(t).
\end{equation}
The response to a forcing with a different temporal evolution can then be computed if its amplitude with respect to the instantaneous doubling forcing is known at each time. In this paper we consider a ramp forcing where the $CO_2$ concentration is increased by 1$\%$ every year until its value is doubled (after about 70 years), and kept constant afterwards. Since the effect of the $CO_2$  increase on the Earth's energy budget scales logarithmically with the concentration, the effect of this forcing on a climate observable can be represented as a ramp function $R_2(t)=\epsilon_{2\times CO_2}R(t)$,  where $R(t)=t/70$ for $t<70$ years, and $R(t)=1$ for $t\ge 70$ years. Since the stabilization value is the same as for the instantaneous doubling experiment $H_2$, the scaling constant is also the same. The  response to the ramp forcing is therefore
\begin{equation}
\delta\Phi_{R_2}(t)\approx\epsilon_{2\times CO_2}\Phi_{R}^{(1)}(t)=\intop_{0}^{t}\epsilon_{2\times CO_2}G_{\Phi}(s)R(t-s)\textrm{d}s.
\end{equation}
Since $\epsilon_{2\times CO_2}G_{\Phi}(t)$ can be computed from data from equation A2, the response  $\delta\Phi_{R_2}(t)$ can be computed even if $\epsilon_{2\times CO_2}$ remains undetermined. This strategy has proved successful in applying linear response theory to climate models of different complexity, including the simulations analysed  in this paper \cite{Ragone2016,Lembo2020}.

When analysing the results of a numerical simulation, we deal with discrete data on a finite time period. Response signals are simply sampled on a finite set of $N+1$ time instants $t_n=n\Delta t$ with $n=0,..,N$, equally spaced by an interval $\Delta t$ over a time domain $[0,T]$, where $T$ is the length of the simulation. The Green's function is similarly sampled on $[0,T-\Delta t]$ at time instants  $t_n=n\Delta t$ with $n=0,..,N-1$, as the derivative of the response signal computed with a standard first order forward approximation
\begin{equation}
\epsilon_{2\times CO_2}G_{\Phi}(t_n)=\frac{\delta\Phi_{H_2}(t_{n+1})-\delta\Phi_{H_2}(t_{n})}{\Delta t}\label{Green_numeric}
\end{equation}
Conversely, the convolution integrals to calculate the response to the $R_2$ scenario from the Green's function are substituted by a sum
\begin{equation}
\delta\Phi_{R_2}(t_n)= \sum_{m=0}^{N-1}\epsilon_{2\times CO_2}G_{\Phi}(t_m)R(t_n-t_m)\Delta t.
\end{equation}
Whilst the discretization of the formulas just presented is  trivial for standard linear response analysis, it becomes less straightforward when dealing with the computation of the proxy response functions and the analysis of causality.

\subsection{Computation of proxy linear response functions from numerical experiments}

The proxy Green's function for a pair of observables is obtained computing the susceptibilities of the Green's functions of the two observables, then computing the proxy susceptibility as their ratio using equation \ref{proxy_susceptibility}, and finally taking the inverse Fourier transform of the proxy susceptibility. The starting point is thus to compute the susceptibilities of the Green's functions taking their Fourier transforms. When working with discrete data on a finite time interval, it is natural to approximate the Fourier transform with the Discrete Fourier Transform (DFT) \cite{Arfken2013}.  However, computing the DFT of the sequence $\epsilon_{2\times CO_2}G_{\Phi}(t_n)$ as defined above would not give a function satysfying the Kramers-Kronig relations \cite{Toll1956,Lucarini2005} and the properties of causality. This issue is discussed in depth in the context of the analysis of spectroscopic data in \cite{BARTHOLDI19739},  where the authors explain how to perform spectral analysis of discrete signals on a finite time interval preserving their causal properties.  Here we follow their approach.

Before computing the Fourier transformation, we extend the sequence on the interval $[-T,T-\Delta t]$ introducing the extended Green's function $\tilde{G}_{\Phi}(t_n)$ for $n=-N,..,N-1$ and
\begin{eqnarray}
\tilde{G}_{\Phi}(t_n)={G}_{\Phi}(t_n),\,\,\,\,\,\,\,\,\,n \ge 0\\
\tilde{G}_{\Phi}(t_n)=0,\,\,\,\,\,\,\,\,\,n < 0
\end{eqnarray}
The susceptibility is then computed up to the constant $\epsilon_{2\times CO_2}$  as the DFT of the extended Green's function on the  set of angular frequencies $\omega_k=2\pi \xi_k$, where $\xi_k= k/(2T)$ is the physical frequency and  $k=-N,...,N-1$
\begin{equation}
\epsilon_{2\times CO_2}\chi_{\Phi}(\omega_k)= \sum_{n=-N}^{N-1} \epsilon_{2\times CO_2}\tilde{G}_{\Phi}(t_n)e^{-i \omega_k t_n}\Delta t.\label{DFT}
\end{equation}
Adding the trail of zeros for negative time to the Green's function enforces causality and guarantees that the DFT of the sequence satisfies the Kramers-Kronig relations. Note that in this way the DFT will have twice as many frequencies than the time steps of the original discrete response signal. This is  equivalent to an interpolation in spectral space and does not alter the information content of the signal \cite{BARTHOLDI19739}.

The proxy susceptibility $\chi_{\Phi_1 \Phi_2}(\omega)$ for a pair of observables $\Phi_1$ and $\Phi_2$ is computed as the ratio of the susceptibilities of predictand and predictor for each discrete angular frequency $\omega_k$
\begin{equation}
\chi_{\Phi_1 \Phi_2}(\omega_k)= \frac{\chi_{\Phi_1}(\omega_k)}{\chi_{\Phi_2}(\omega_k)}
\end{equation}
where the scaling constant has disappeared since it is the same for both observables. The proxy Green's function is finally computed  as the inverse Fourier transform of the proxy susceptibility, here  approximated by the inverse DFT
\begin{equation}
G_{\Phi_1 \Phi_2}(t_n)= \sum_{k=-N}^{N-1} \chi_{\Phi_1 \Phi_2}(\omega_k)e^{ i \omega_k t_n}\Delta \omega  \label{DFT}
\end{equation}
for each $t_n$ in $[-T,T-\Delta t]$ and with $\Delta \omega=1/(2T)$.

\subsection{Computation of analytic continuation of proxy susceptibility from numerical experiments}
When the proxy Green's function is a causal function and the proxy susceptibility admits analytic continuation $\chi_{\Phi_1 \Phi_2}(\sigma)$ in the upper complex plane, this can be obtained numerically by  computing the discrete Laplace transform of the proxy Green's function.  Defining a set of complex frequencies $\sigma_{j_k}=\lambda_j+i\omega_k$, we have
\begin{equation}
\chi_{\Phi_1 \Phi_2}(\lambda_j+i \omega_k)= \sum_{n=-N}^{N-1} G_{\Phi_1 \Phi_2}(t_n)e^{(\lambda_j+i \omega_k)t_n}\Delta t.\label{discrete_continuation}
\end{equation}
where the discrete rates $\lambda_j$ can be chosen arbitrarily and in our case are  $\lambda_j= j/(2T)$, for $j=0,...,N-1$. This operation  is well defined only when the proxy Green's function is a causal function. When this is not the case, the true proxy susceptibility will have one or more singularities in the upper complex plane. For discrete complex frequencies $\sigma_{jk}$ that are close  enough to a singularity of the true proxy susceptibility, the numerical proxy susceptibility will take extremely large values that will make \ref{discrete_continuation}  numerically untreatable. Note that, because of the integral nature of the analytic continuation, in this case the computation of the proxy susceptibility will break down not only in the neighborhoods of the singularities, but for any complex frequency beyond the radius of the singularity closest to zero. 

In order to identify the location of the singularities of the proxy susceptibility we can exploit the fact that in our case this is the ratio of two functions that we know admit analytic continuation. The analytic continuation of a ratio, if it exists, is the ratio of the analytic continuations of numerator and denominator. The singularities of the proxy susceptibility can thus only be at complex frequencies where  the denominator (the susceptibility of the predictor) is zero, whilst the numerator (the susceptibility of the predictand) is different from zero.

We can therefore consider the ratio of the discrete Laplace transforms of the extended Green's functions of predictand and predictor
\begin{equation}
\chi^{test}_{\Phi_1 \Phi_2}(\lambda_j+i \omega_k)=\frac{ \sum_{n=-N}^{N-1}\epsilon_{2\times CO_2} \tilde{G}_{\Phi_1}(t_n)e^{(\lambda_j+i \omega_k)t_n}\Delta t}{\sum_{n=-N}^{N-1}\epsilon_{2\times CO_2}\tilde{G}_{\Phi_2}(t_n)e^{(\lambda_j+i \omega_k)t_n}\Delta t}
\end{equation}
If $G_{\Phi_1 \Phi_2}(t_n)$ is a causal function, then $\chi_{\Phi_1 \Phi_2}(\lambda_j+i \omega_k)=\chi^{test}_{\Phi_1 \Phi_2}(\lambda_j+i \omega_k)$ and so this ratio will effectively give the proxy susceptibility. If $G_{\Phi_1 \Phi_2}(t)$ is not causal then $\chi^{test}_{\Phi_1 \Phi_2}(\lambda_j+i \omega_k)$ will still be computable except than at the singularities. Since the numerical susceptibility ad denominator will not have in general true zeroes, but just very small values for discrete frequencies near to the true zero, it will be possible to identify the frequencies responsible for the singularity by looking for pronounced local minima of the module of $\chi^{test}_{\Phi_1 \Phi_2}(\lambda_j+i \omega_k)$, or, as done in the main text, pronounced local maxima of the module of its reciprocal $\chi^{test}_{\Phi_2 \Phi_1}(\lambda_j+i \omega_k)$.


\vskip2pc










\bibliographystyle{RS} 

\bibliography{RSPA_Author_tex} 

@PREAMBLE{
 "\providecommand{\noopsort}[1]{}" 
 # "\providecommand{\singleletter}[1]{#1}%" 
}

@misc{data2xCO2abrupt,
      url = {https://doi.org/10.26050/WDCC/2xCO2abrupt},
      title = {Beyond Forcing Scenarios: Predicting Climate Change through Response Operators in a Coupled General Circulation Model: CO2 abrupt doubling experiment},
      publisher = {World Data Center for Climate (WDCC) at DKRZ},
      year = {2022},
      author = {Lunkeit, Frank and Lembo, Valerio and Lucarini, Valerio},
      doi = {10.26050/WDCC/2xCO2abrupt}
}

@misc{data1pctCO2,
      url = {https://doi.org/10.26050/WDCC/1pctCO2},
      title = {Beyond Forcing Scenarios: Predicting Climate Change through Response Operators in a Coupled General Circulation Model: 1\% annual CO2 increase ramp experiment},
      publisher = {World Data Center for Climate (WDCC) at DKRZ},
      year = {2022},
      author = {Lembo, Valerio and Lunkeit, Frank and Lucarini, Valerio},
      doi = {10.26050/WDCC/1pctCO2}
}

@article{PIRES2024133988,
title = {A general theory to estimate Information transfer in nonlinear systems},
journal = {Physica D: Nonlinear Phenomena},
volume = {458},
pages = {133988},
year = {2024},
issn = {0167-2789},
doi = {https://doi.org/10.1016/j.physd.2023.133988},
url = {https://www.sciencedirect.com/science/article/pii/S0167278923003421},
author = {Carlos A. Pires and David Docquier and Stéphane Vannitsem}
}

@article{Liang2016,
  title = {Information flow and causality as rigorous notions ab initio},
  author = {Liang, X. San},
  journal = {Phys. Rev. E},
  volume = {94},
  issue = {5},
  pages = {052201},
  numpages = {28},
  year = {2016},
  month = {Nov},
  publisher = {American Physical Society},
  doi = {10.1103/PhysRevE.94.052201},
  url = {https://link.aps.org/doi/10.1103/PhysRevE.94.052201}
}

@article{Hall2019,
	author = {Hall, Alex and Cox, Peter and Huntingford, Chris and Klein, Stephen},
	date = {2019/04/01},
	doi = {10.1038/s41558-019-0436-6},
	id = {Hall2019},
	isbn = {1758-6798},
	journal = {Nature Climate Change},
	number = {4},
	pages = {269--278},
	title = {Progressing emergent constraints on future climate change},
	url = {https://doi.org/10.1038/s41558-019-0436-6},
	volume = {9},
	year = {2019}}

@article{VAROTSOS2025106556,
title = {Emergent constraints for uncertainty reduction in climate projections},
journal = {Journal of Atmospheric and Solar-Terrestrial Physics},
volume = {274},
pages = {106556},
year = {2025},
issn = {1364-6826},
doi = {https://doi.org/10.1016/j.jastp.2025.106556},
url = {https://www.sciencedirect.com/science/article/pii/S1364682625001403},
author = {C. Varotsos and M. Efstathiou and N. Sarlis}
}

@article{Shiogama2022,
	author = {Shiogama, Hideo and Watanabe, Masahiro and Kim, Hyungjun and Hirota, Nagio},
	date = {2022/02/01},
	doi = {10.1038/s41586-021-04310-8},
	id = {Shiogama2022},
	isbn = {1476-4687},
	journal = {Nature},
	number = {7898},
	pages = {612--616},
	title = {Emergent constraints on future precipitation changes},
	url = {https://doi.org/10.1038/s41586-021-04310-8},
	volume = {602},
	year = {2022}}

@article{BARTHOLDI19739,
title = {Fourier spectroscopy and the causality principle},
journal = {Journal of Magnetic Resonance (1969)},
volume = {11},
number = {1},
pages = {9-19},
year = {1973},
issn = {0022-2364},
doi = {https://doi.org/10.1016/0022-2364(73)90076-0},
url = {https://www.sciencedirect.com/science/article/pii/0022236473900760},
author = {E Bartholdi and R.R Ernst}
}

@article{BodaiLucarini2020Chaos,
    author = {Bódai, Tamás and Lucarini, Valerio and Lunkeit, Frank},
    title = {Can we use linear response theory to assess geoengineering strategies?},
    journal = {Chaos: An Interdisciplinary Journal of Nonlinear Science},
    volume = {30},
    number = {2},
    pages = {023124},
    year = {2020},
    month = {02},
    issn = {1054-1500},
    doi = {10.1063/1.5122255},
    url = {https://doi.org/10.1063/1.5122255},
    eprint = {https://pubs.aip.org/aip/cha/article-pdf/doi/10.1063/1.5122255/14625276/023124_1_online.pdf},
}

@book{Arfken2013,
  title={Mathematical Methods for Physicists: A Comprehensive Guide},
  author={Arfken, George B. and Weber, Hans J. and Harris, Frank E.},
  edition={7},
  year={2013},
  publisher={Academic Press},
  address={Boston},
  isbn={978-0-12-384654-9}
}

@article{Toll1956,
  title = {Causality and the Dispersion Relation: Logical Foundations},
  author = {Toll, John S.},
  journal = {Phys. Rev.},
  volume = {104},
  issue = {6},
  pages = {1760--1770},
  year = {1956},
  month = {Dec},
  publisher = {American Physical Society},
  doi = {10.1103/PhysRev.104.1760},
  url = {https://link.aps.org/doi/10.1103/PhysRev.104.1760}
}

@book{Lucarini2005,
  address = {Berlin, Heidelberg},
  series = {Springer Series in Optical Sciences},
  title = {Kramers-{Kronig} {Relations} in {Optical} {Materials} {Research}},
  volume = {110},
  isbn = {978-3-540-23673-3 978-3-540-27316-5},
  url = {https://link.springer.com/book/10.1007/b138913},
  language = {en},
  publisher = {Springer Berlin Heidelberg},
  author = {Lucarini, Valerio and Saarinen, Jarkko J. and Peiponen, Kai-Erik and Vartiainen, Erik M.},
  year = {2005},
  doi = {10.1007/b138913},
}

@article{Deltatto2025,
    author = {Del Tatto, Vittorio and Banerjee, Debarshi and Hassanali, Ali and Laio, Alessandro},
    title = {Towards a robust approach to infer causality from molecular dynamics simulations},
    journal = {The Journal of Chemical Physics},
    volume = {162},
    number = {24},
    pages = {244105},
    year = {2025},
    month = {06},
    issn = {0021-9606},
    doi = {10.1063/5.0267926},
    url = {https://doi.org/10.1063/5.0267926},
    eprint = {https://pubs.aip.org/aip/jcp/article-pdf/doi/10.1063/5.0267926/20567438/244105_1_5.0267926.pdf},
}

@article{Allione2025,
  title = {Linear Scaling Causal Discovery from High-Dimensional Time Series by Dynamical Community Detection},
  author = {Allione, Matteo and Del Tatto, Vittorio and Laio, Alessandro},
  journal = {Phys. Rev. Lett.},
  volume = {135},
  issue = {4},
  pages = {047401},
  numpages = {8},
  year = {2025},
  month = {Jul},
  publisher = {American Physical Society},
  doi = {10.1103/kd73-93cg},
  url = {https://link.aps.org/doi/10.1103/kd73-93cg}
}

@book{Pearl2009,
  author    = {Judea Pearl},
  title     = {Causality: Models, Reasoning, and Inference},
  edition   = {2nd},
  year      = {2009},
  publisher = {Cambridge University Press},
  address   = {New York}
}

@article{Granger1969,
  author = {Granger, C. W. J.},
  title = {Investigating causal relations by econometric models and cross-spectral methods},
  journal = {Econometrica},
  volume = {37},
  number = {3},
  pages = {424--438},
  year = {1969},
  publisher = {The Econometric Society}
}

@inbook{Pearl1995,
  author    = {J. Pearl},
  title     = {From Bayesian networks to causal networks},
  booktitle = {Mathematical Models for Handling Partial Knowledge in Artificial Intelligence},
  pages={157-182},
  publisher = {Springer},
  address   = {Boston},
  year      = {1995}
}

@book{Peixoto,
    author ={Peixoto, J.P. and Oort, A.H}, 
    title = {Physics of Climate},
    publisher = {American Institute of Physics, New York},
    year ={1992}
}

@article{Pendergrass2014,
  author    = {Pendergrass, Angeline G. and Hartmann, Dennis L.},
  title     = {Changes in the Distribution of Rain Frequency and Intensity
               in Response to Global Warming},
  journal   = {Journal of Climate},
  volume    = {27},
  number    = {22},
  pages     = {8372--8383},
  year      = {2014},
  doi       = {10.1175/JCLI-D-14-00183.1},
  url       = {https://doi.org/10.1175/JCLI-D-14-00183.1},
}

@article{Haerter2009,
  author    = {Haerter, Jan O. and Berg, Peter},
  title     = {Unexpected rise in extreme precipitation caused by a shift in rain type?},
  journal   = {Nature Geoscience},
  volume    = {2},
  number    = {6},
  pages     = {372--373},
  year      = {2009},
  doi       = {10.1038/ngeo523},
  url       = {https://doi.org/10.1038/ngeo523},
}

@article{Berg2013,
  author = {Berg, Peter and Moseley, Christopher and Haerter, Jan O.},
  title = {Strong increase in convective precipitation in response to higher temperatures},
  journal = {Nature Geoscience},
  year = {2013},
  volume = {6},
  number = {3},
  pages = {181--185},
  doi = {10.1038/ngeo1731},
  publisher = {Nature Publishing Group}}

@article{HeldSoden2006,
      author = "Isaac M. Held and Brian J. Soden",
      title = "Robust Responses of the Hydrological Cycle to Global Warming",
      journal = "Journal of Climate",
      year = "2006",
      publisher = "American Meteorological Society",
      address = "Boston MA, USA",
      volume = "19",
      number = "21",
      doi = "10.1175/JCLI3990.1",
      pages=      "5686 - 5699",
      url = "https://journals.ametsoc.org/view/journals/clim/19/21/jcli3990.1.xml"
}

@article{Lucarini2025,
      title={Interpretable and Equation-Free Response Theory for Complex Systems}, 
      author={Valerio Lucarini},
      year={2025},
      journal={Phil. Trans. Roy. Soc. A},
      doi={10.1098/rsta.2025.0081},
      url = {https://arxiv.org/abs/2502.07908}
}

@book{Titchmarsh1939,
	author = {Titchmarsh, E. C.},
	title = {The theory of functions /},
	publisher = {Oxford university press,},
	year = {1939.},
	address = {[London]},
	edition = {2d ed.}
}

@article{Zaglietal2026,
author = {Zagli, Niccol\`{o} and Colbrook, Matthew J. and Lucarini, Valerio and Mezi\'{c}, Igor and Moroney, John},
title = {Bridging the Gap Between Koopmanism and Response Theory: Using Natural Variability to Predict Forced Response},
journal = {SIAM Journal on Applied Dynamical Systems},
volume = {25},
number = {1},
pages = {196-229},
year = {2026},
doi = {10.1137/24M1699206},

URL = { 
    
        https://doi.org/10.1137/24M1699206
    
    

},
eprint = { 
    
        https://doi.org/10.1137/24M1699206
    
    

}
}

@article{abramov2007,
	Author = {Abramov, R.V. and Majda, A.J.},
	Journal = {Nonlinearity},
	Number = {12},
	Pages = {2793},
	Title = {Blended response algorithms for linear fluctuation-dissipation for complex nonlinear dynamical systems},
	Volume = {20},
	Year = {2007}}

@article {Cooper2011,
      author = "Fenwick C. Cooper and Peter H. Haynes",
      title = "Climate Sensitivity via a Nonparametric Fluctuation–Dissipation Theorem",
      journal = "Journal of the Atmospheric Sciences",
      year = "2011",
      publisher = "American Meteorological Society",
      address = "Boston MA, USA",
      volume = "68",
      number = "5",
      doi = "10.1175/2010JAS3633.1",
      pages=      "937 - 953",
      url = "https://journals.ametsoc.org/view/journals/atsc/68/5/2010jas3633.1.xml"
}

@article{giorgini2024linear,
  title   = {Response Theory via Generative Score Modeling},
  author  = {Giorgini, Ludovico T. and Deck, Katherine and Bischoff, Tobias and Souza, Andre N.},
  journal = {Physical Review Letters},
  volume  = {133},
  number  = {26},
  pages   = {267302},
  year    = {2024},
  doi     = {10.1103/PhysRevLett.133.267302}
}

@article{giorgini2024datadriven,
  title   = {Predicting forced responses of probability distributions via the fluctuation--dissipation theorem and generative modeling},
  author  = {Giorgini, Ludovico T. and Falasca, Fabrizio and Souza, Andre N.},
  journal = {Proceedings of the National Academy of Sciences},
  volume  = {122},
  number  = {41},
  pages   = {e2509578122},
  year    = {2025},
  doi     = {10.1073/pnas.2509578122}
}

@article{vonderHeydt2016,
	Author = {von der Heydt, Anna S. and Dijkstra, Henk A. and van de Wal, Roderik S. W. and Caballero, Rodrigo and Crucifix, Michel and Foster, Gavin L. and Huber, Matthew and K{\"o}hler, Peter and Rohling, Eelco and Valdes, Paul J. and Ashwin, Peter and Bathiany, Sebastian and Berends, Tijn and van Bree, Loes G. J. and Ditlevsen, Peter and Ghil, Michael and Haywood, Alan M. and Katzav, Joel and Lohmann, Gerrit and Lohmann, Johannes and Lucarini, Valerio and Marzocchi, Alice and P{\"a}like, Heiko and Baroni, Itzel Ruvalcaba and Simon, Dirk and Sluijs, Appy and Stap, Lennert B. and Tantet, Alexis and Viebahn, Jan and Ziegler, Martin},
	Day = {01},
	Doi = {10.1007/s40641-016-0049-3},
	Issn = {2198-6061},
	Journal = {Current Climate Change Reports},
	Month = dec,
	Number = {4},
	Pages = {148--158},
	Title = {Lessons on Climate Sensitivity From Past Climate Changes},
	Volume = {2},
	Year = {2016}}

@article{ruelle_nonequilibrium_1998,
	Author = {Ruelle, D.},
	Journal = {Nonlinearity},
	Month = jan,
	Number = {1},
	Pages = {5--18},
	Shorttitle = {Nonequilibrium statistical mechanics near equilibrium},
	Title = {Nonequilibrium statistical mechanics near equilibrium: computing higher-order terms},
	Volume = {11},
	Year = {1998}}

@article{Lucarini2008,
  title   = {Response Theory for Equilibrium and Non-Equilibrium Statistical Mechanics: Causality and Generalized {K}ramers-{K}ronig relations},
  author  = {Lucarini, Valerio},
  journal = {Journal of Statistical Physics},
  volume  = {131},
  number  = {3},
  pages   = {543--558},
  year    = {2008},
  doi     = {10.1007/s10955-008-9498-y}
}

@book{imkeller2001,
  title     = {Stochastic Climate Models},
  editor    = {Imkeller, Peter and Von Storch, Jin-Song},
  series    = {Progress in Probability},
  volume    = {49},
  year      = {2001},
  publisher = {Birkhäuser Basel},
  address   = {Basel, Boston},
  isbn      = {978-3-7643-6520-2},
  doi       = {10.1007/978-3-0348-8287-3},
  language  = {english},
  pages     = {xxvii, 398}
}

@article{Lucarinietal2017,
	Author = {Lucarini, Valerio and Ragone, Francesco and Lunkeit, Frank},
	Da = {2017/02/01},
	Doi = {10.1007/s10955-016-1506-z},
	Id = {Lucarini2017},
	Isbn = {1572-9613},
	Journal = {J. Stat. Phys.},
	Number = {3},
	Pages = {1036--1064},
	Title = {Predicting Climate Change Using Response Theory: Global Averages and Spatial Patterns},
	Ty = {JOUR},
	Url = {https://doi.org/10.1007/s10955-016-1506-z},
	Volume = {166},
	Year = {2017}}

@article{Tel2020,
   author = {Tél, T. and Bódai, T. and Drótos, G. and Haszpra, T. and Herein, M. and Kaszás, B. and Vincze, M.},
   title = {The Theory of Parallel Climate Realizations},
   journal = {Journal of Statistical Physics},
   volume = {179},
   number = {5},
   pages = {1496-1530},
   ISSN = {1572-9613},
   DOI = {10.1007/s10955-019-02445-7},
   url = {https://doi.org/10.1007/s10955-019-02445-7},
   year = {2020},
   type = {Journal Article}
}

@article{Bodai2013,
  title = {Driving a conceptual model climate by different processes: Snapshot attractors and extreme events},
  author = {B\'odai, Tam\'as and K\'arolyi, Gy\"orgy and T\'el, Tam\'as},
  journal = {Phys. Rev. E},
  volume = {87},
  issue = {2},
  pages = {022822},
  numpages = {10},
  year = {2013},
  month = {Feb},
  publisher = {American Physical Society},
  doi = {10.1103/PhysRevE.87.022822},
  url = {https://link.aps.org/doi/10.1103/PhysRevE.87.022822}
}

@article{Maher2021,
AUTHOR = {Maher, N. and Milinski, S. and Ludwig, R.},
TITLE = {Large ensemble climate model simulations: introduction, overview, and future prospects for utilising multiple types of large ensemble},
JOURNAL = {Earth System Dynamics},
VOLUME = {12},
YEAR = {2021},
NUMBER = {2},
PAGES = {401--418},
URL = {https://esd.copernicus.org/articles/12/401/2021/},
DOI = {10.5194/esd-12-401-2021}
}

@article{Mauritsen2019,
author = {Mauritsen, Thorsten and Bader, Jürgen and Becker, Tobias and Behrens, Jörg and Bittner, Matthias and Brokopf, Renate and Brovkin, Victor and Claussen, Martin and Crueger, Traute and Esch, Monika and Fast, Irina and Fiedler, Stephanie and Fläschner, Dagmar and Gayler, Veronika and Giorgetta, Marco and Goll, Daniel S. and Haak, Helmuth and Hagemann, Stefan and Hedemann, Christopher and Hohenegger, Cathy and Ilyina, Tatiana and Jahns, Thomas and Jimenéz-de-la-Cuesta, Diego and Jungclaus, Johann and Kleinen, Thomas and Kloster, Silvia and Kracher, Daniela and Kinne, Stefan and Kleberg, Deike and Lasslop, Gitta and Kornblueh, Luis and Marotzke, Jochem and Matei, Daniela and Meraner, Katharina and Mikolajewicz, Uwe and Modali, Kameswarrao and Möbis, Benjamin and Müller, Wolfgang A. and Nabel, Julia E. M. S. and Nam, Christine C. W. and Notz, Dirk and Nyawira, Sarah-Sylvia and Paulsen, Hanna and Peters, Karsten and Pincus, Robert and Pohlmann, Holger and Pongratz, Julia and Popp, Max and Raddatz, Thomas Jürgen and Rast, Sebastian and Redler, Rene and Reick, Christian H. and Rohrschneider, Tim and Schemann, Vera and Schmidt, Hauke and Schnur, Reiner and Schulzweida, Uwe and Six, Katharina D. and Stein, Lukas and Stemmler, Irene and Stevens, Bjorn and von Storch, Jin-Song and Tian, Fangxing and Voigt, Aiko and Vrese, Philipp and Wieners, Karl-Hermann and Wilkenskjeld, Stiig and Winkler, Alexander and Roeckner, Erich},
title = {Developments in the MPI-M Earth System Model version 1.2 (MPI-ESM1.2) and Its Response to Increasing CO2},
journal = {Journal of Advances in Modeling Earth Systems},
volume = {11},
number = {4},
pages = {998-1038},
doi = {https://doi.org/10.1029/2018MS001400},
url = {https://agupubs.onlinelibrary.wiley.com/doi/abs/10.1029/2018MS001400},
eprint = {https://agupubs.onlinelibrary.wiley.com/doi/pdf/10.1029/2018MS001400},
year = {2019}
}

@Article{Eyring2016,
AUTHOR = {Eyring, V. and Bony, S. and Meehl, G. A. and Senior, C. A. and Stevens, B. and Stouffer, R. J. and Taylor, K. E.},
TITLE = {Overview of the Coupled Model Intercomparison Project Phase 6 (CMIP6)
experimental design and organization},
JOURNAL = {Geoscientific Model Development},
VOLUME = {9},
YEAR = {2016},
NUMBER = {5},
PAGES = {1937--1958},
URL = {https://gmd.copernicus.org/articles/9/1937/2016/},
DOI = {10.5194/gmd-9-1937-2016}
}

@article{Leith1975,
	Author = {Leith, C. E.},
	Journal = {J. Atmos. Sci.},
	Pages = {2022},
	Title = {Climate response and fluctuation dissipation},
	Volume = {32},
	Year = {1975}}

@article{Santos2022,
doi = {10.1088/1751-8121/ac90fd},
url = {https://dx.doi.org/10.1088/1751-8121/ac90fd},
year = {2022},
month = {oct},
publisher = {IOP Publishing},
volume = {55},
number = {42},
pages = {425002},
author = {Manuel Santos Guti\'errez and Valerio Lucarini},
title = {On some aspects of the response to stochastic and deterministic forcings},
journal = {Journal of Physics A: Mathematical and Theoretical}
}

@article{Lucarinietal2026,
title = {A general framework for linking free and forced fluctuations via Koopmanism},
journal = {Chaos, Solitons and Fractals},
volume = {202},
pages = {117540},
year = {2026},
issn = {0960-0779},
doi = {https://doi.org/10.1016/j.chaos.2025.117540},
url = {https://www.sciencedirect.com/science/article/pii/S096007792501553X},
author = {Valerio Lucarini and Manuel Santos Gutiérrez and John Moroney and Niccolò Zagli}
}

@article{HairerMajda2010,
	Author = {Hairer, M. and Majda, A. J.},
	Coden = {NONLE5},
	Doi = {10.1088/0951-7715/23/4/008},
	Fjournal = {Nonlinearity},
	Issn = {0951-7715},
	Journal = {Nonlinearity},
	Mrclass = {82C05 (37A50 37A60 37N20 60H10)},
	Mrnumber = {2602020 (2011d:82048)},
	Mrreviewer = {Da-Quan Jiang},
	Number = {4},
	Pages = {909--922},
	Title = {A simple framework to justify linear response theory},
	Url = {http://dx.doi.org/10.1088/0951-7715/23/4/008},
	Volume = {23},
	Year = {2010}}

@article{Bettolo2008,
	Author = {U. Marini Bettolo Marconi and A. Puglisi and L. Rondoni and A. Vulpiani},
	Journal = {Phys. Rep.},
	Pages = {111},
	Title = {Fluctuation-Dissipation: Response Theory in Statistical Physics},
	Volume = 461,
	Year = {2008}}

@article{Sarracino2019,
	Author = {Sarracino, A. and Vulpiani, A.},
	Journal = {Chaos},
	Pages = {083132},
	Title = {On the fluctuation-dissipation relation in non-equilibrium and non-Hamiltonian systems},
	Volume = {29},
	Year = {2019}}

@article{Asbrink2023,
    author = {Asbrink, Leif},
    title = {Revisiting causality using stochastics on atmospheric temperature and CO2 concentration},
    journal = {Proceedings of the Royal Society A: Mathematical, Physical and Engineering Sciences},
    volume = {479},
    number = {2269},
    pages = {20220529},
    year = {2023},
    month = {01},
    issn = {1364-5021},
    doi = {10.1098/rspa.2022.0529},
    url = {https://doi.org/10.1098/rspa.2022.0529},
    eprint = {https://royalsocietypublishing.org/rspa/article-pdf/doi/10.1098/rspa.2022.0529/727622/rspa.2022.0529.pdf},
}

@article {Bracegirdle2012,
      author = "Thomas J. Bracegirdle and David B. Stephenson",
      title = "On the Robustness of Emergent Constraints Used in Multimodel Climate Change Projections of Arctic Warming",
      journal = "Journal of Climate",
      year = "2012",
      publisher = "American Meteorological Society",
      address = "Boston MA, USA",
      volume = "26",
      number = "2",
      doi = "10.1175/JCLI-D-12-00537.1",
      pages=      "669 - 678",
      url = "https://journals.ametsoc.org/view/journals/clim/26/2/jcli-d-12-00537.1.xml"
}

@article{Cox2018,
  title = {Emergent constraint on equilibrium climate sensitivity from global temperature variability},
  author = {Cox, PM and Huntingford, C and Williamson MS},
  journal = {Nature},
  volume = {553},
  issue = {7688},
  pages = {319-322},
  year = {2018}
  }

@article{GhilLucarini2020,
  title = {The physics of climate variability and climate change},
  author = {Ghil, Michael and Lucarini, Valerio},
  journal = {Rev. Mod. Phys.},
  volume = {92},
  issue = {3},
  pages = {035002},
  numpages = {77},
  year = {2020},
  month = {Jul},
  publisher = {American Physical Society},
  doi = {10.1103/RevModPhys.92.035002},
  url = {https://link.aps.org/doi/10.1103/RevModPhys.92.035002}
}

@article{Gordon2014,
author = {Gordon, Neil D. and Klein, Stephen A.},
title = {Low-cloud optical depth feedback in climate models},
journal = {Journal of Geophysical Research: Atmospheres},
volume = {119},
number = {10},
pages = {6052-6065},
doi = {https://doi.org/10.1002/2013JD021052},
url = {https://agupubs.onlinelibrary.wiley.com/doi/abs/10.1002/2013JD021052},
eprint = {https://agupubs.onlinelibrary.wiley.com/doi/pdf/10.1002/2013JD021052},
year = {2014}
}

@article{Hall2006,
author = {Hall, Alex and Qu, Xin},
title = {Using the current seasonal cycle to constrain snow albedo feedback in future climate change},
journal = {Geophysical Research Letters},
volume = {33},
number = {3},
pages = {},
doi = {https://doi.org/10.1029/2005GL025127},
url = {https://agupubs.onlinelibrary.wiley.com/doi/abs/10.1029/2005GL025127},
eprint = {https://agupubs.onlinelibrary.wiley.com/doi/pdf/10.1029/2005GL025127},
year = {2006}
}

@book{IPCC_2021_WGI,
  address = {Cambridge, UK and New York, NY, USA},
  doi = {10.1017/9781009157896},
  editor = {Masson-Delmotte, V. and Zhai, P. and Pirani, A. and Connors, S. L. and Péan, C. and Berger, S. and Caud, N. and Chen, Y. and Goldfarb, L. and Gomis, M. I. and Huang, M. and Leitzell, K. and Lonnoy, E. and Matthews, J. B. R. and Maycock, T. K. and Waterfield, T. and Yelekçi, O. and Yu, R. and Zhou, B.},
  pages = {2391},
  publisher = {Cambridge University Press},
  title = {Climate Change 2021: The Physical Science Basis. Contribution of Working Group I to the Sixth Assessment Report of the Intergovernmental Panel on Climate Change},
  type = {Book},
  url = {https://report.ipcc.ch/ar6/wg1/IPCC_AR6_WGI_FullReport.pdf},
  year = {2021}
}

@article{Klein2015,
	author = {Klein, Stephen A. and Hall, Alex},
	date = {2015/12/01},
	doi = {10.1007/s40641-015-0027-1},
	id = {Klein2015},
	isbn = {2198-6061},
	journal = {Current Climate Change Reports},
	number = {4},
	pages = {276--287},
	title = {Emergent Constraints for Cloud Feedbacks},
	url = {https://doi.org/10.1007/s40641-015-0027-1},
	volume = {1},
	year = {2015}}

@article {Knutti2006,
      author = "Reto Knutti and Gerald A. Meehl and Myles R. Allen and David A. Stainforth",
      title = "Constraining Climate Sensitivity from the Seasonal Cycle in Surface Temperature",
      journal = "Journal of Climate",
      year = "2006",
      publisher = "American Meteorological Society",
      address = "Boston MA, USA",
      volume = "19",
      number = "17",
      doi = "10.1175/JCLI3865.1",
      pages=      "4224 - 4233",
      url = "https://journals.ametsoc.org/view/journals/clim/19/17/jcli3865.1.xml"
}

@article{Koutsoyiannis2022a,
    author = {Koutsoyiannis, Demetris and Onof, Christian and Christofides, Antonis and Kundzewicz, Zbigniew W.},
    title = {Revisiting causality using stochastics: 1. Theory},
    journal = {Proceedings of the Royal Society A: Mathematical, Physical and Engineering Sciences},
    volume = {478},
    number = {2261},
    pages = {20210835},
    year = {2022},
    issn = {1364-5021},
    doi = {10.1098/rspa.2021.0835},
    url = {https://doi.org/10.1098/rspa.2021.0835},
    eprint = {https://royalsocietypublishing.org/rspa/article-pdf/doi/10.1098/rspa.2021.0835/727146/rspa.2021.0835.pdf},
}

@article{Koutsoyiannis2022b,
    author = {Koutsoyiannis, Demetris and Onof, Christian and Christofidis, Antonis and Kundzewicz, Zbigniew W.},
    title = {Revisiting causality using stochastics: 2. Applications},
    journal = {Proceedings of the Royal Society A: Mathematical, Physical and Engineering Sciences},
    volume = {478},
    number = {2261},
    pages = {20210836},
    year = {2022},
    doi = {10.1098/rspa.2021.0836},
    url = {https://doi.org/10.1098/rspa.2021.0836},
    eprint = {https://royalsocietypublishing.org/rspa/article-pdf/doi/10.1098/rspa.2021.0836/727338/rspa.2021.0836.pdf},
}

@article{Lembo2020,
	author = {Lembo, Valerio and Lucarini, Valerio and Ragone, Francesco},
	date = {2020/05/26},
	doi = {10.1038/s41598-020-65297-2},
	id = {Lembo2020},
	isbn = {2045-2322},
	journal = {Scientific Reports},
	number = {1},
	pages = {8668},
	title = {Beyond Forcing Scenarios: Predicting Climate Change through Response Operators in a Coupled General Circulation Model},
	url = {https://doi.org/10.1038/s41598-020-65297-2},
	volume = {10},
	year = {2020}}

@article{Lucarini2018,
	author = {Lucarini, Valerio},
	date = {2018/12/01},
	doi = {10.1007/s10955-018-2151-5},
	id = {Lucarini2018},
	isbn = {1572-9613},
	journal = {Journal of Statistical Physics},
	number = {6},
	pages = {1698--1721},
	title = {Revising and Extending the Linear Response Theory for Statistical Mechanical Systems: Evaluating Observables as Predictors and Predictands},
	url = {https://doi.org/10.1007/s10955-018-2151-5},
	volume = {173},
	year = {2018}}

@Article{Nijsse2018,
AUTHOR = {Nijsse, F. J. M. M. and Dijkstra, H. A.},
TITLE = {A mathematical approach to understanding emergent constraints},
JOURNAL = {Earth System Dynamics},
VOLUME = {9},
YEAR = {2018},
NUMBER = {3},
PAGES = {999--1012},
URL = {https://esd.copernicus.org/articles/9/999/2018/},
DOI = {10.5194/esd-9-999-2018}
}

@Article{Nijsse2020,
AUTHOR = {Nijsse, F. J. M. M. and Cox, P. M. and Williamson, M. S.},
TITLE = {Emergent constraints on transient climate response (TCR) and equilibrium climate sensitivity (ECS) from historical warming in CMIP5 and CMIP6 models},
JOURNAL = {Earth System Dynamics},
VOLUME = {11},
YEAR = {2020},
NUMBER = {3},
PAGES = {737--750},
URL = {https://esd.copernicus.org/articles/11/737/2020/},
DOI = {10.5194/esd-11-737-2020}
}

@article {Nobre2023,
      author = "Nobre, P. and Veiga, S.F. and Giarolla, E. et al.",
      title = "AMOC decline and recovery in a warmer climate",
      journal = "Sci Rep",
      year = "2023",
      volume = "13",
      number = "15928",
      doi = "https://doi.org/10.1038/s41598-023-43143-5",
      pages=      "15928"
}

@Article{Nowack2025,
AUTHOR = {Nowack, P. and Watson-Parris, D.},
TITLE = {Opinion: Why all emergent constraints are wrong but some are useful -- a machine learning perspective},
JOURNAL = {Atmospheric Chemistry and Physics},
VOLUME = {25},
YEAR = {2025},
NUMBER = {4},
PAGES = {2365--2384},
URL = {https://acp.copernicus.org/articles/25/2365/2025/},
DOI = {10.5194/acp-25-2365-2025}
}

@article{Qu2014,
	author = {Qu, Xin and Hall, Alex},
	date = {2014/01/01},
	doi = {10.1007/s00382-013-1774-0},
	id = {Qu2014},
	isbn = {1432-0894},
	journal = {Climate Dynamics},
	number = {1},
	pages = {69--81},
	title = {On the persistent spread in snow-albedo feedback},
	url = {https://doi.org/10.1007/s00382-013-1774-0},
	volume = {42},
	year = {2014}}

@article{Ragone2016,
	author = {Ragone, Francesco and Lucarini, Valerio and Lunkeit, Frank},
	journal = {Climate Dynamics},
	number = {5},
	pages = {1459--1471},
	title = {A new framework for climate sensitivity and prediction: a modelling perspective},
	volume = {46},
	year = {2016}}

@Article{Ruelle1998,
AUTHOR = {Ruelle, D.},
TITLE = {General linear response formula in statistical mechanics, and
the fluctuation-dissipation theorem far from equilibrium},
JOURNAL = {Physics Letters A},
VOLUME = {245},
YEAR = {1998},
NUMBER = {3-4},
PAGES = {220–224},
}

@Article{Ruelle2009,
AUTHOR = {Ruelle, D.},
TITLE = {A review of linear response theory for general differentiable dynamical systems},
JOURNAL = {Nonlinearity},
VOLUME = {22},
YEAR = {2009},
NUMBER = {4},
PAGES = {855–870},
}

@article {Stephens2008,
      author = "Graeme L. Stephens and Todd D. Ellis",
      title = "Controls of Global-Mean Precipitation Increases in Global Warming GCM Experiments",
      journal = "Journal of Climate",
      year = "2008",
      publisher = "American Meteorological Society",
      address = "Boston MA, USA",
      volume = "21",
      number = "23",
      doi = "10.1175/2008JCLI2144.1",
      pages=      "6141 - 6155",
      url = "https://journals.ametsoc.org/view/journals/clim/21/23/2008jcli2144.1.xml"
}

@article{Terhaar2020,
	author = {Terhaar, Jens and Kwiatkowski, Lester and Bopp, Laurent},
	date = {2020/06/01},
	doi = {10.1038/s41586-020-2360-3},
	id = {Terhaar2020},
	isbn = {1476-4687},
	journal = {Nature},
	number = {7812},
	pages = {379--383},
	title = {Emergent constraint on Arctic Ocean acidification in the twenty-first century},
	url = {https://doi.org/10.1038/s41586-020-2360-3},
	volume = {582},
	year = {2020}}

@article{Tomasini2021,
	author = {Tomasini, Umberto Maria and Lucarini, Valerio},
	date = {2021/10/01},
	doi = {10.1140/epjs/s11734-021-00158-1},
	id = {Tomasini2021},
	isbn = {1951-6401},
	journal = {The European Physical Journal Special Topics},
	number = {14},
	pages = {2813--2832},
	title = {Predictors and predictands of linear response in spatially extended systems},
	url = {https://doi.org/10.1140/epjs/s11734-021-00158-1},
	volume = {230},
	year = {2021}}

@article{Wenzel2014,
author = {Wenzel, Sabrina and Cox, Peter M. and Eyring, Veronika and Friedlingstein, Pierre},
title = {Emergent constraints on climate-carbon cycle feedbacks in the CMIP5 Earth system models},
journal = {Journal of Geophysical Research: Biogeosciences},
volume = {119},
number = {5},
pages = {794-807},
doi = {https://doi.org/10.1002/2013JG002591},
url = {https://agupubs.onlinelibrary.wiley.com/doi/abs/10.1002/2013JG002591},
eprint = {https://agupubs.onlinelibrary.wiley.com/doi/pdf/10.1002/2013JG002591},
year = {2014}
}

@article{Williamson2021,
  title = {Emergent constraints on climate sensitivities},
  author = {Williamson, Mark S. and Thackeray, Chad W. and Cox, Peter M. and Hall, Alex and Huntingford, Chris and Nijsse, Femke J. M. M.},
  journal = {Rev. Mod. Phys.},
  volume = {93},
  issue = {2},
  pages = {025004},
  numpages = {34},
  year = {2021},
  month = {May},
  publisher = {American Physical Society},
  doi = {10.1103/RevModPhys.93.025004},
  url = {https://link.aps.org/doi/10.1103/RevModPhys.93.025004}
}

@article{Zappa2020,
author = {Giuseppe Zappa  and Paulo Ceppi  and Theodore G. Shepherd },
title = {Time-evolving sea-surface warming patterns modulate the climate change response of subtropical precipitation over land},
journal = {Proceedings of the National Academy of Sciences},
volume = {117},
number = {9},
pages = {4539-4545},
year = {2020},
doi = {10.1073/pnas.1911015117},
URL = {https://www.pnas.org/doi/abs/10.1073/pnas.1911015117},
eprint = {https://www.pnas.org/doi/pdf/10.1073/pnas.1911015117},
}

@article{hasselmann1976,
	author = {Hasselmann, K.},
	doi = {10.3402/tellusa.v28i6.11316},
	issn = {0040-2826},
	journal = {Tellus},
	number = {6},
	pages = {473--485},
	title = {{Stochastic climate models Part I. Theory}},
	volume = {28},
	year = {1976}}

@article{LucariniChekroun2024,
  title = {Detecting and Attributing Change in Climate and Complex Systems: Foundations, Green's Functions, and Nonlinear Fingerprints},
  author = {Lucarini, Valerio and Chekroun, Micka\"el D.},
  journal = {Phys. Rev. Lett.},
  volume = {133},
  issue = {24},
  pages = {244201},
  numpages = {11},
  year = {2024},
  month = {Dec},
  publisher = {American Physical Society},
  doi = {10.1103/PhysRevLett.133.244201},
  url = {https://link.aps.org/doi/10.1103/PhysRevLett.133.244201}
}

@article{LucariniChekroun2023,
	author = {Lucarini, Valerio and Chekroun, Micka{\"e}l D.},
	date = {2023/12/01},
	doi = {10.1038/s42254-023-00650-8},
	id = {Lucarini2023},
	isbn = {2522-5820},
	journal = {Nature Reviews Physics},
	number = {12},
	pages = {744--765},
	title = {Theoretical tools for understanding the climate crisis from Hasselmann's programme and beyond},
	url = {https://doi.org/10.1038/s42254-023-00650-8},
	volume = {5},
	year = {2023}}

\end{document}